\documentclass[useAMS,usenatbib]{mnras}
\usepackage{graphicx}

\usepackage{caption}
\usepackage{amsmath}
\usepackage{natbib}
\usepackage{color}
\usepackage[english]{babel}

\def \aj{AJ}
\def \aap{A\&A}
\def \apj{ApJ}
\def \apjs{ApJS}
\def \mnras{MNRAS}

\def \jcap{JCAP}
\def \jkas{JKAS}
\def \na{NewAst}
\def \nat{Nature}

\def \ascl Astrophysics Source Code Library
\def \pasj{PASJ}

\newcommand{\noop}[1]{}

\title[Cosmological gas in simulations]{Resolution Convergence in Cosmological Hydrodynamical Simulations Using Adaptive
Mesh Refinement }
\author[Snaith et al.]{Owain N. Snaith$^{1}$, Changbom Park$^{1}$, Juhan Kim$^{2}$\thanks{Corresponding author: kjhan@kias.re.kr}, Joakim Rosdahl$^{3}$\\
$^1$ School of Physics, Korea Institute for Advanced Study, 85 Hoegiro, Dongdaemungu, Seoul 02455, Korea.\\
$^2$ Center for Advanced Computation, Korea Institute for Advanced Study, 85 Hoegiro, Dongdaemungu, Seoul 02455, Korea.\\
$^3$Univ. Lyon, Univ. Lyon 1, Ens de Lyon, CNRS, Centre de Recherche Astrophysique de Lyon UMR5574, F-69230 Saint-Genis-Laval, France\\
}
\date{\today}
\begin{document}

\maketitle

\begin{abstract}
We have explored the evolution of gas distributions from cosmological simulations carried out using the \textsc{RAMSES} adaptive mesh refinement (AMR) code, to explore the effects of resolution on cosmological hydrodynamical simulations. It is vital to understand the effect of both the resolution of initial conditions and the final resolution of the simulation. Lower initial resolution simulations tend to produce smaller numbers of low mass structures. This will strongly affect the assembly history of objects, and has the same effect of simulating different cosmologies. 
The resolution of initial conditions is an important factor in simulations, even with a fixed maximum spatial resolution. The power spectrum of gas in simulations using AMR diverges strongly from the fixed grid approach - with more power on small scales in the AMR simulations - even at fixed physical resolution and also produces offsets in the star formation at specific epochs. This is because before certain times the upper grid levels are held back to maintain approximately fixed physical resolution, and to mimic the natural evolution of dark matter only simulations.
Although the impact of hold back falls with increasing spatial and initial-condition resolutions, the offsets in the star formation remain down to a spatial resolution of 1 kpc.  
These offsets are of order of 10-20\%, which is below the uncertainty in the implemented physics but are expected to affect the detailed properties of galaxies. We have implemented a new grid-hold-back approach to minimize the impact of hold back on the star formation rate.

\end{abstract}

\section{Introduction}  
Cosmological simulations are valuable tools for improving our understanding of the formation of structures in the Universe. For many years, large pure dark matter cosmological simulations have been capable of modelling large-scale properties of the Universe \citep[e.g.][]{Kim2015, Springel2005,Riebe2013,Teyssier2009,Kim2011,Kim2009,Park1990}. Recently, however, improved methods and computing resources have allowed the development of large-scale hydrodynamical cosmological simulations \citep{Vogelsberger2014,Nelson2015, Schaye2015, Dubois2014,Crain2015,Gottloeber2010}. 

A principal advantage of such simulations is that they can be compared to observations from large redshift surveys of cosmic objects \citep{Li2016,2016,Choi2010,Park2005,Park1994}. However, the predictions of these simulations depend on various factors, such as the properties of the initial conditions, the simulation resolution, and the implemented sub-grid physics. These have been explored extensively in a number of works, such as \citet{Kim2014}, \citet{Frenk1999}, \citet{LHuillier2014}, \citet{Klypin2017}, \citet{Schneider2016} and \citet{Rudd2008}, among others.

Various factors in the initial conditions (ICs) affect the final outcome of cosmological simulations. For example, \citet{LHuillier2014} use N-body simulations carried out using \textsc{Gadget-III} \citep{Springel2005b, Gadgetsoft} and \textsc{Gotpm} \citep{Dubinski2004,Park2005}, analysing the effect of the starting redshift, the order of the perturbation theory used for the initial conditions, the simulation code, and whether the pre-ICs are glass or gridded, while \citet{Rudd2008} studied the effect of baryons and cooling on the matter power spectrum. Such studies are vital to the future of cosmology because current surveys are reaching the point where the matter power spectrum of the Universe can be defined to very fine precision \citep[e.g.][]{Dawson2013, Huterer2005}. However, various authors have shown that on small scales N-body simulations are unable to precisely model the matter power spectrum, because of baryonic effects \citep{vanDaalen2011,Schneider2015,Schneider2016,Rudd2008}. 

The density field of the $\Lambda$CDM universe contains structure on many different scales. In simulations, this is limited by both the simulation volume, which defines the maximum scale, and the simulation resolution which sets the minimum meaningful scale. The simulation resolution is commonly given as the gravitational softening length (in particle codes) or the minimum cell size (in grid codes). However, in such cases, the initial power is truncated at the resolution of the initial conditions, which can be several orders of magnitude less than the force resolution of the {\it simulation}. \citet{Schneider2016} note that AMR codes require higher resolution than SPH codes at early times, in order to produce similar small scale power. This difference decreases at later times.

Cosmological probes have recently become incredibly precise with the advent of missions such as {\it \textsc{Planck}} \citep[e.g.][]{Ade2016}, SDSS \citep{Albareti2016} or WiggleZ \citep{Drinkwater2010}, for example. Thus, the results of simulations are being tested like never before. The exquisite observations require ever more precise simulations, meaning that approximations which have previously been acceptable are no longer adequate \citep{Huterer2005}.

Many simulation papers quote their resolution as the force resolution of the simulation runs, i.e. the highest resolution. However, these scales are often orders of magnitude smaller than the resolution of the initial conditions. It is the initial conditions, however, that define the minimum scale length of the initial fluctuations in density and/or velocity, as well as the mass resolution of the dark matter component. It is critically important to check whether or not the simulation is representative of the intended cosmological model, given the box size of the simulation, and the mass resolution of the initial conditions. Thus, when comparing simulations one must be mindful of the initial conditions resolution, because information and degrees of freedom can be missing, even if the quoted force resolution is the same. 

In this paper we explore the effect of resolution on the outcome of the gas component of cosmological simulations, and disentangle the contrasting effects of the simulations resolution (force resolution) and the initial conditions resolution (IC resolution). The effect of various aspects of the simulation on the dark matter (or total matter, which is dominated by the dark matter) component has been extensively studied previously \citep[e.g.][]{Schneider2016,LHuillier2014}.

\citet{Teyssier2002} showed that for a given set of initial conditions the small scale power increases along with the simulation force resolution, \citep[see Figure 11 of ][]{Teyssier2002}. In \textsc{RAMSES} this is controlled by the number of refinement levels. Furthermore, \citet{Rasera2006} note that the mass resolution, which is set by the resolution of the initial conditions, has a strong impact on the smallest structures which form, and the distribution of star formation. This is because halos with fewer than 400 particles will not form stars in their simulation. This means that the distribution of stars and feedback will be different as the force and initial conditions resolution are changed. This will also impact the power spectrum of the gas due to feedback effects. 

 Below the scale on which we can readily calculate the power spectrum of the simulation we must rely on other diagnostics. Small-scale fluctuations of the gas influence, and are effected by, the star formation physics. Thus, for small scales, in high resolution simulations, we utilize the cosmic star formation rate as a complementary measure of the gas properties.

We will introduce the initial conditions and simulation code in section \ref{Sec:Method} and then present the results of our analysis in section \ref{Sec:Results} using the results of \textsc{RAMSES} simulations with different resolutions  and present a potential solution to some of the points made in section \ref{Sec:New}. In section \ref{Sec:Conclusions} we discuss our findings and present our conclusions.

\section{Methods}
\label{Sec:Method}
\subsection{Simulations}

\subsubsection{Initial Conditions}
We generate initial conditions using \textsc{Music}\footnote{https://bitbucket.org/ohahn/music} \citep{Hahn2011,Musicsoft}. \textsc{Music}
is designed to produce Gaussian random fields from a given power spectrum (in our case a $\Lambda$CDM power spectrum). The method relies on Lagrangian Perturbation Theory (LPT), which is implemented up to second order terms (LPT2). The code generates Gaussian initial conditions based on a random number seed provided by the user, which means that the initial conditions are entirely reproducible, and different resolution simulations could have the same configurations of density fluctuations at the overlapped frequencies. 

\textsc{Music} uses a real-space transfer function which is periodically replicated along the length of the box. This differs from other codes which sample in $k-$space \citep[e.g.][]{Fraschetti2010,Bertschinger2001}. Ultimately, however, the end result is the same to very high accuracy \citep{Hahn2011}.

We use \textsc{Music} to generate both dark matter and baryonic initial conditions, which are generalized in the code as a two-phase fluid. We use the second order mode in LPT when calculating the initial velocity and density fields, for improved accuracy in higher order modes \citep{LHuillier2014}. We utilize a transfer function using the fitting formula of \citet{Eisenstein1998}, which includes baryon acoustic oscillations. Using this initial power spectrum the velocity and density fields for the dark matter and the gas are the same.

\begin{figure}
\centering
\includegraphics[scale=.55]{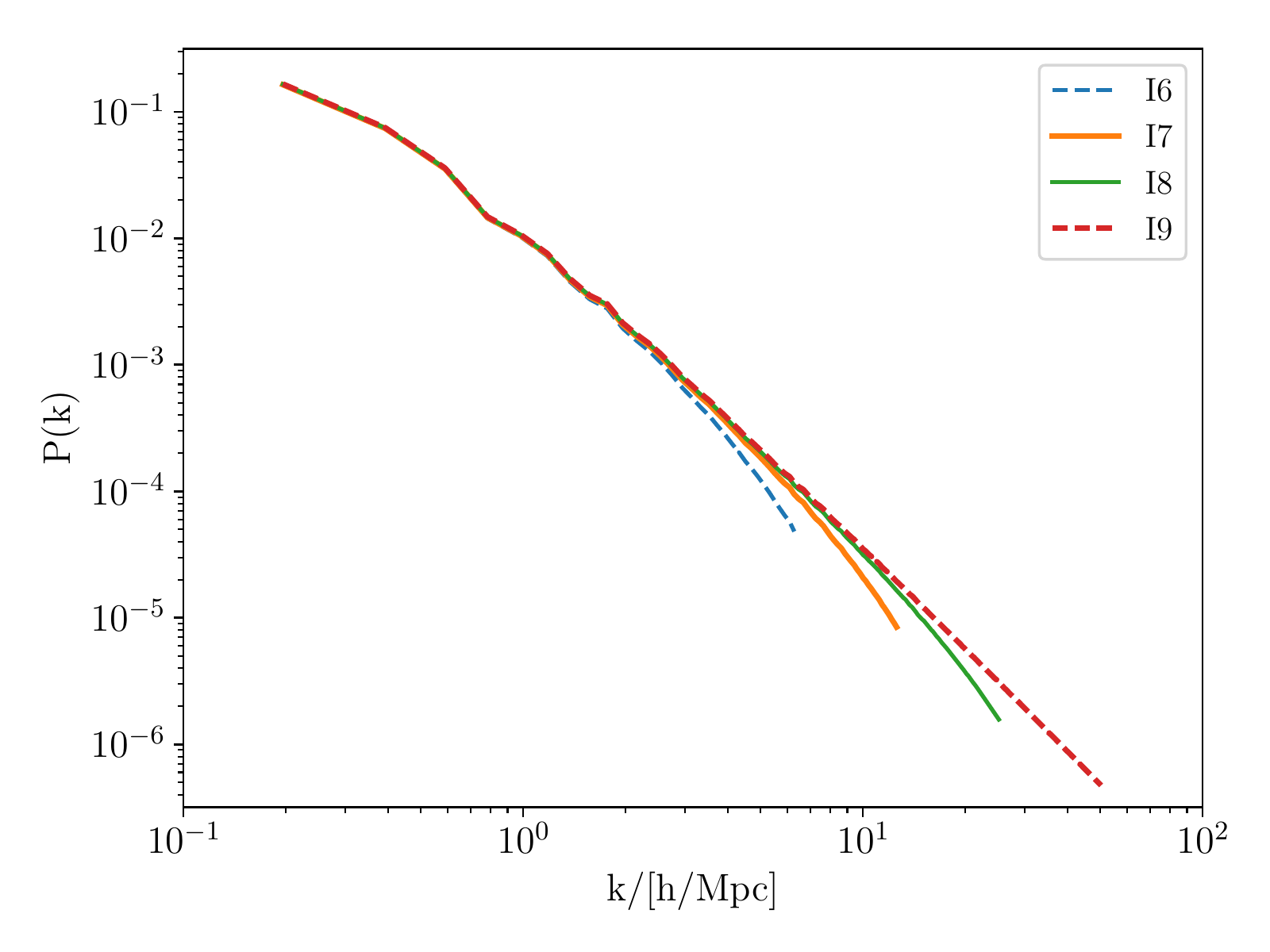} 
\caption{The power spectrum of gas at $z$=100 for each of the I6Fy, I7Fy, I8Fy and I9Fy simulations used in this paper, where $y$ is the maximum refinement level to me chosen (see Table \ref{Tab:DeltaNS}). The lines are coloured by the IC resolution as shown in the legend.  $k$ is the wavenumber and $P(k)$ is the power at a given $k$.}
\label{Fig:Initial_Pk}
\end{figure}

We set the initial redshift to be $z$=100, and use a cosmology of ($\mathrm{\Omega_m}$, $\mathrm{\Omega_\lambda}$, $\mathrm{\Omega_b}$, $\mathrm{H_0}$, $\sigma_8$, n$_{\mathrm{spec}}$) = (0.276, 0.724, 0.045, 70.3, 0.811, 0.961) in a 32 $h^{-1}$Mpc box \citep[e.g.][]{Hahn2014}. The initial power spectra are shown in Fig. \ref{Fig:Initial_Pk}. These initial conditions are used throughout the paper, and will be indicated by the prefix `Ix' throughout, where x is 6, 7, 8 or 9. This will be referred to as the IC resolution. Furthermore, we will characterise the simulation resolution, the formal resolution of the simulation, as a second `Fy' where y is the maximum refinement level of the simulation. The simulation will be quoted as {\it ``IxFy''} in the  text. The length scale being probed by a given level is given as 

\begin{equation}
S = \frac{L_{B}}{2^{l}}
\end{equation}
where $L_{B}$ is the box size and $l$ is the grid level. For the initial conditions $l$=$x$, and the maximum possible resolution has $l$=$y$.

As the initial conditions resolution increases, more small-scale fluctuations are encoded in the initial power spectrum as shown in Fig. \ref{Fig:Initial_Pk}. For example, the I9Fy initial conditions contain information between  wavenumbers $k$=6~$h$/Mpc and $k$=50~$h$/Mpc which is not present in the I6Fy initial conditions. In each simulation we use the same set of random number seeds. This means that the overlapped frequencies in the initial power spectrum are identical, i.e. there is no cosmic variance to take into account between simulations. Additionally, the simulation resolution sets the minimum spacial scale, while the IC resolution sets the dark matter mass resolution. This means that the I6Fy, I7Fy, I8Fy and I9Fy simulations have a dark matter particle mass of 1$\times$10$^{10} $M$_\odot$, 1.4$\times$10$^{9} $M$_\odot$, 1.8$\times$10$^{8} $M$_\odot$ and 2.2$\times$10$^{7} $M$_\odot$ respectively. As the gas mass is adaptive, the minimum gas mass is  limited by the simulation resolution or $l_{\mathrm{fine}}$. The stellar mass is set by the simulation resolution as well. \citet{Schneider2016} note that accurate (to within a few percent) power spectra are attained to $k$=10 $h$/Mpc for particle masses 1$\times$10$^{9} $M$_\odot$.

\subsubsection{RAMSES}

We use the Adaptive Mesh Refinement (AMR)  
 code called \textsc{RAMSES} \citep{Teyssier2002,Ramsessoft}, which follows the evolution of dark matter, gas and stars across cosmic time. We use the standard publicly available version of \textsc{RAMSES} in  most of this paper\footnote{https://bitbucket.org/rteyssie/ramses/ The  commit of \textsc{RAMSES} on bitbucket is for the version used in section 3 is 3b581a77e113f32c62036eafa6bd640ddbee88a4}. \textsc{RAMSES} follows dark matter, gas, and stars using an adaptive mesh. Dark matter is traced by particles, using gravitational forces calculated on the mesh, while the hydrodynamical component is discretized on the mesh, rather than using particles. This means that the mass resolution of the dark matter is set by the initial conditions, while the gas resolution is refined into smaller and smaller units until the simulation resolution is attained.   

Refinement can be carried out in various ways, but we chose the Quasi-Lagrangian method typical of cosmological simulations. Using this approach, the refinements are triggered by the number of particles in a cell, as well as a given gas density threshold. 

However, when star formation is allowed, \textsc{RAMSES} includes a second criterion for refinement. In order to maintain approximately constant force resolution in physical units the upper levels of refinement are ``held back'' so that the volume can be refined to maintain the physical resolution \citep{Dubois2014}. This prevents unjustifiably high physical force resolution at early times, as will be discussed in detail later. In \textsc{RAMSES}, the resolution is intrinsically comoving because the simulation volume is subdivide by the mesh, and the entire volume grows as the universe expands. Many of the physical processes, however, rely on constant physical force scales and this must be taken into account via the hold-back approach. 

The criteria for star formation are based on the temperature of the gas, a density threshold, and a Kennikutt-Schmidt law \citep{Kennicutt1998,Dubois2012}. Star formation is formally deactivated by, for example, setting the star formation efficiency to zero. Stars, once formed, return energy into the ISM from supernovae  within around 10 Myr, which act to heat the gas and prevent overcooling \citep{Dubois2008}. 

One interesting aspect of AMR simulations is that AMR codes are less effective at resolving halos with lower particle numbers. Thus, AMR simulations are more expensive than treecodes because they require higher resolutions in order to resolve the same mass of object at early times \citep{Schneider2016}. This greater sensitivity to resolution is inherent in the AMR methodology rather than the \textsc{RAMSES} code itself, and its effect is especially prevalent at high redshift \citep[e.g.][]{OShea2005}. This is important if one wishes to compare simulations carried out using different simulation codes. For example, in  \citet{Schneider2016} a comparison between 
\textsc{RAMSES} with a spacial resolution of 2 $h^{-1} $kpc (and an IC resolution of 244 $h^{-1}$kpc) and  \textsc{Gadget3} with a gravitational softening length of 10~$h^{-1}$kpc (using the same ICs) resulted in a matter power spectrum in \textsc{RAMSES} having a 10\% offset at $z$=3.8 for $k\sim$10 $h^{-1}$Mpc compared to \textsc{Gadget3}. This offset decreased over time as larger halos dominate the power spectrum \citep{Schneider2016}. For a further code comparison and tests of different hydrosolvers on a galaxy scale also see \citet{Few2016},  \citet{Schneider2016}, \citet{OShea2005}, \citet{Hahn2011} among others.

We do not tune the subgrid physics of the simulation to any particular observations,  because the purpose of this paper is to explore how the properties of the simulation change with resolution. For the star formation physics we use $\epsilon_{star}$=0.02, $\Delta_{star}$=55, $n_{star}$=0.1, $m_{star}$=0.2, $\eta_{sn}$=0.1 for the fully star forming runs, where $\epsilon_{star}$, $\Delta_{star}$, $n_{star}$, $m_{star}$, $\eta_{sn}$ are the star formation efficiency, the star formation threshold in critical density, the star formation threshold in $H/cm^{-3}$, the star particle mass in units of the SPH mass, and the supernova mass fraction respectively. The star formation efficiency is the ratio of the free fall time and the star formation time scale or the efficiency with which gas is transformed into stars.  $m_{star}$ is defined according to the SPH mass, which is given by $mass_{SPH} = \Omega_b/\Omega_m 2^{-(3x)}$, where $x$ is the grid level of the initial conditions and $\Omega_b$ and $\Omega_m$ are the baryon and total matter fractions. The supernova mass fraction is the fraction of a stars mass that is ejected when it undergoes a supernova. We use the basic supernova feedback implementation in \textsc{RAMSES}, which is simple thermal feedback. 

In order to turn off star formation we set $\epsilon_{star}$=0.0. This, however, also turns off the grid hold-back in our version of \textsc{RAMSES}. We set $n_{\rm star}=10^6$ to allow grid hold back but prevent star formation. This last parameter also means, however, that the polytropic pressure support is turned off. The polytropic pressure support sets a minimum temperature for the gas of a given density and acts to ensure that the Jeans length is at least 4 grid cells \citep{Teyssier2013,Gabor2016}. The polytropic temperature used is 10,000K. This does not strongly affect the power spectrum compared to the star formation physics or the impact of the grid hold-back, at least for the $k$ values of interest in this paper. The other parameters are kept constant, in order to assess the impact of the resolution as the only varying parameter.

\subsection{Calculating the Power spectrum}

The power spectrum of matter in the simulation is calculated on a 512$^3$ grid (an effective resolution of 62.5 $h^{-1}$kpc) with each grid cell containing

\begin{equation}
\delta_{i,j,k} \equiv \frac{\rho_{i,j,k}}{\bar{\rho}_{\text{gas,dm}}}-1,
\end{equation}
\noindent
where $i$, $j$ and $k$ are the index of the cell from 1 to 512,  $\rho$ is the density of a cell, and $\bar{\rho}$ is the mean density of the gas, dark matter or total matter. Unless otherwise stated we use the same species on the denominator as the numerator rather than total matter. The only expected difference is the precise normalisation of the power spectrum rather than its shape, which is of primary interest here. The power spectrum is then,

\begin{equation}
P(\mathbf{k}) = \langle |{\delta(\mathbf{k})}|^2 \rangle,
\end{equation}

\begin{equation}
\delta(\mathbf{k}) = \frac{1}{V}\int \delta(\mathbf{x})\exp(i \mathbf{k}\cdot \mathbf{x})d^3\mathbf{x},
\label{Eqn:delta}
\end{equation}
\noindent
where V is the volume of the simulation box, $\mathbf{k}$ is the Fourier mode and $\mathbf{x}$ is the position.  $k$ is defined as $2\pi m /V^{1/3}$, where $m$ is the mode number from 1 to 512. 
 
We calculate the power spectrum on a 512$^3$ grid, which corresponds to a level 9 grid. This corresponds to a resolution of 62.5 $h^{-1}$kpc, and is used throughout the paper. Since we are using data from the entire periodic volume, no window function is required.

As we are sampling the density field at certain points in the gas, we do not need to correct the recovered power spectrum, as in \citet{Cui2008}. However, the finite cell size explains the deviation in the different initial power spectra in Fig. \ref{Fig:Initial_Pk} near the Nyquist frequency. For the dark matter, we assign mass to the regular grid using a cloud-in-cell approach and so the corresponding correction is required \citep{Cui2008}.

\section{Results}    
\label{Sec:Results}
\begin{table}
\centering
\begin{tabular}{lccccc}
\hline

Name & l$_{\mathrm{min}}$ & l$_{\mathrm{max}}$ & Resolution\\
& & & $h^{-1}$kpc\\
\hline
I6F9 & 6 & 9 &62.5\\
I7F9 & 7 & 9 &62.5\\
I8F9 & 8 & 9 &62.5\\
I9F9 & 9 & 9 &62.5\\
\hline
I6F15 & 6 & 15 &1$^*$ \\
I7F15 & 7 & 15 &1 \\
I8F15 & 8 & 15 &1 \\
\hline
I7F10 & 7 & 10 &32.25 \\
I7F12 & 7 & 12 &8 \\
I7F14 & 7 & 14 &2 \\
\hline
\end{tabular}
\caption{Summary of the simulations used in this paper. $^*$ this corresponds to 1.4 kpc, which is the gravitational softening length used in Illustris-2 \citep{Vogelsberger2014}.}
\label{Tab:DeltaNS}
\end{table}

\subsection{Comparison with a fixed grid simulation}

In order to explore the impact of IC resolution on the results of cosmological simulations, we compare a set of simulations with different initial resolutions but the same maximum force resolution. To understand the impact of the AMR methodology implemented in \textsc{RAMSES} we also run a static grid simulation at the same force resolution. This requirement limits the available force resolution due to the extensive resource requirement of static grid simulations with high resolution. In \S \ref{Sec:Default} we use simulations with initial coarse grids of 6, 7, 8 and 9, corresponding to resolutions of 500, 250, 125 and 62.5 $h^{-1}$kpc respectively, and a force resolution of 62.5 $h^{-1}$kpc. Thus, the level 9 simulation (I9F9) is the fixed grid one, and is used as a reference.  These simulations are summarised in Table \ref{Tab:DeltaNS}.

\subsubsection{Default runs with star formation (IxF9 simulations)}
\label{Sec:Default}

We expect that the power spectra of lower mass resolution simulations will converge towards the power spectrum of the highest mass resolution run, in this case I9F9 as the resolution in the initial conditions increases. This is the goal of most numerical convergence studies, that the results of a simulation converge to a given answer as the simulation becomes increasingly precise. In this case, we expect to see this convergent behaviour because more of the $\Lambda$CDM power spectrum is traced as the IC resolution increases, and there are more degrees of freedom. We also expect that the difference between I9F9 and the other runs will decrease {\it smoothly}. 

\begin{figure}
\centering
\includegraphics[scale=.5]{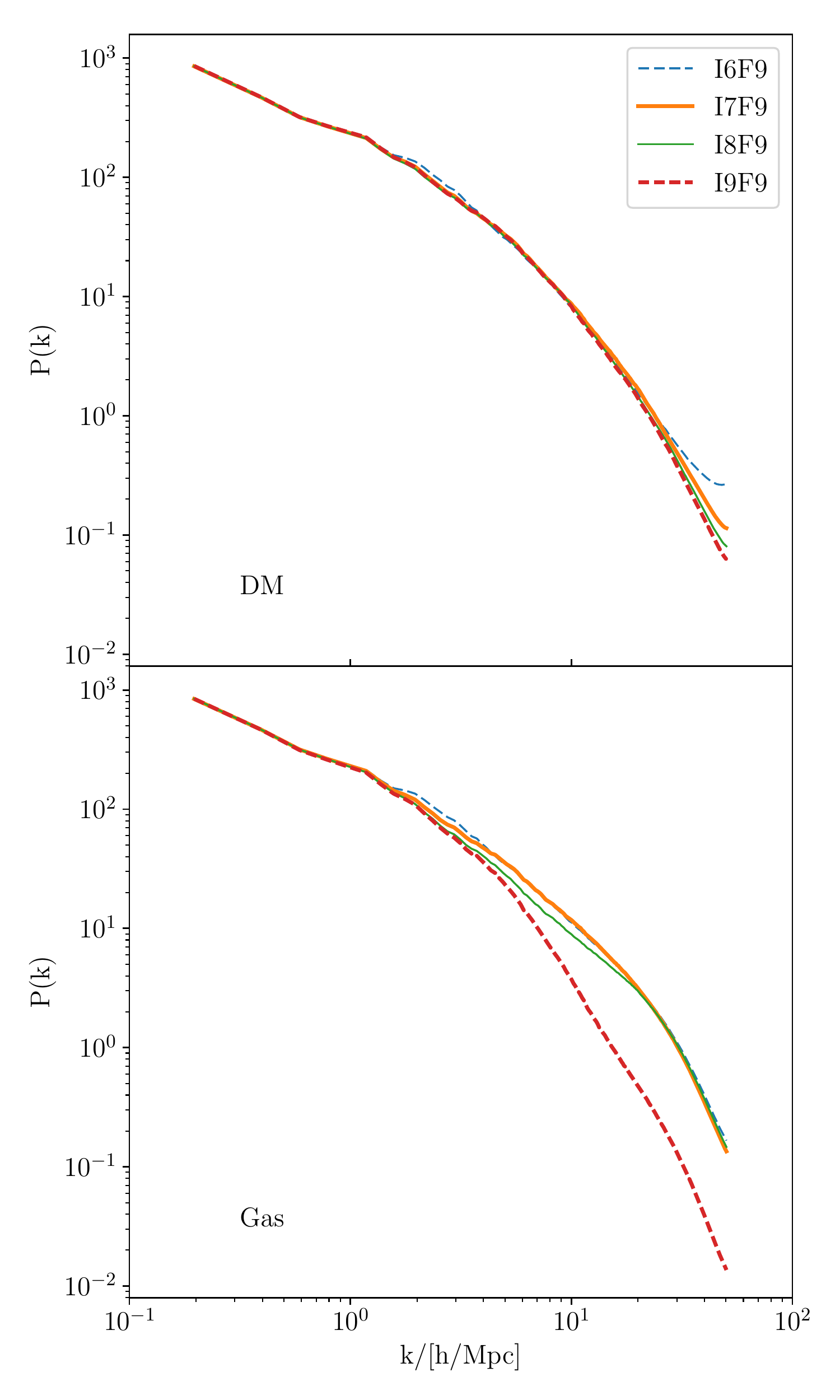} 
\caption{Power spectra for dark matter (top panel) and gas (bottom panel) at $z$=0 for different initial resolutions. The final simulation resolution is fixed 512 cells in each dimension, which corresponds to a spatial scale of 62.5 $h^{-1}$kpc.  The initial conditions used here are shown in Fig. \ref{Fig:Initial_Pk} with the same colours.  }
\label{Fig:Figure1}
\end{figure}

The upper panel of Fig. \ref{Fig:Figure1} illustrates that the $z$=0 power spectrum of the {\it dark matter} converges smoothly towards the high-resolution instance. The difference between the various AMR simulations and the reference simulation decreases with increasing resolution. The dark matter component varies smoothly, with the greatest difference at the Nyquist frequency ($\sim$50.46 $h$/Mpc), as expected. The {\it gas} power spectrum (lower panel), however, shows a large jump in the small-scale power between the AMR cases and the reference simulation, and there is no convergence at all. Furthermore, this difference occurs across a much larger range of $k$ than in the dark matter, with divergence from approximately 5.0 $h$/Mpc, rather than just close to the Nyquist frequency. 

\begin{figure}
\centering
\includegraphics[scale=.5]{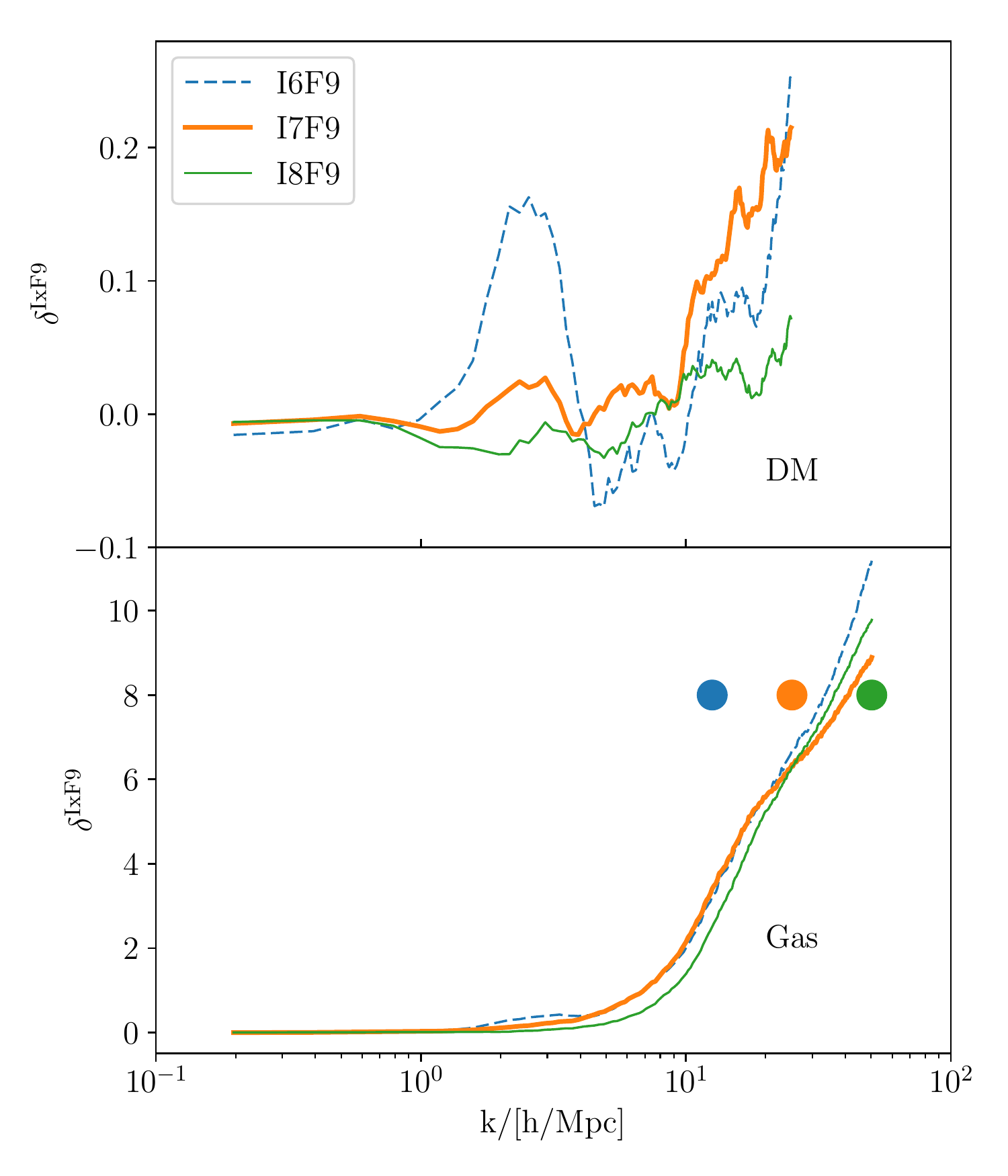} 
\caption{Differences of the power amplitude ($\delta^{\mathrm{IxF9}}_{\mathrm{I9F9}}$) between each simulation (I6F9, I7F9 and I8F9) and the reference simulation (I9F9) with the same colours as shown in Fig. \ref{Fig:Figure1}. The top panel is for the dark matter and the bottom panel is for the gas. The filled circles in the bottom panel are illustrative of the $k$ value of the refinement levels 7, 8 and 9 painted in blue, orange and green, respectively. These values are given by $k=(2\pi/32 [Mpc/h]) 2^{\mathrm{level}-1}$, where the volume is (32 h$^{-1}$Mpc)$^3$}
\label{Fig:Figure1b}
\end{figure}

Figure \ref{Fig:Figure1b} shows the differences between the AMR runs and the reference simulation, defined by

\begin{equation}
\delta^{\mathrm{sim}}_{\mathrm{ref}} \equiv \frac{P(k)_{\mathrm{sim}}}{P(k)_{\mathrm{ref}}}-1,
\end{equation}
\noindent
where $P(k)_{\mathrm{sim}}$ is the power spectrum of a given simulation, and $P(k)_{\mathrm{ref}}$ is the reference run \citep[e.g.][]{LHuillier2014,Cui2008}. For the dark matter, the difference only exceeds 1 for the I6F9 run  and then only near the Nyquist frequency. For I7F9 and I8F9 the differences at $\log(k)$=1.5 are $\delta^{\text{IxF9}}=0.8$ and $0.3$ respectively. The gas component is, by far, more concerning. The difference in small scale power reaches as high as $\delta^{\mathrm{IxF9}}=12.0$. Furthermore, the difference between the AMR runs is much smaller than the difference between the AMR runs and the reference. The differences between I6F9 (I7F9) and I8F9 is 0.3 (-0.07). This implies a considerable divergence between simulations carried out using a fixed grid and an adaptive mesh.

One potential reason for this disagreement is that the initial conditions introduce more small scale power at higher resolution, i.e. the initial power spectrum extends to higher values of $k$. However, although the specifics of the power spectrum show some differences caused by the loss of initial small scale power, the major difference is in the AMR versus fixed grid runs rather than between different AMR runs. \citet{OShea2005} compared an AMR simulation at high redshift with one carried out using \textsc{Gadget II} \citep{Springel2005}. The match between the codes was poor at early times before the density exceeded the threshold for refinement in the AMR code, but improved over time (from $z$=10 to $z$=3) but their results do not extend to $z$=0.

As the $z$=0 gas power spectra in runs with AMR diverge so significantly from fixed-grid simulations, it is vital to ascertain why. If this difference is not understood it becomes difficult to understand the results of cosmological AMR simulations generally. If the gas distribution shows a strong dependency on the choice of simulation approach, even on comparatively large scales (e.g. $k$=125 $h^{-1}$kpc, Fig. \ref{Fig:Figure1b}), identifying which simulation results are scientifically useful is difficult.

As the initial conditions show much greater similarity than the gas power spectra at $z$=0, this difference clearly develops over the run of the simulation. Thus, understanding the evolution of the power spectrum becomes critical in interpreting the difference between the AMR and reference simulations. 

\begin{figure*}
\centering
\includegraphics[scale=.55,trim={.cm .0cm .0cm 0cm},clip]{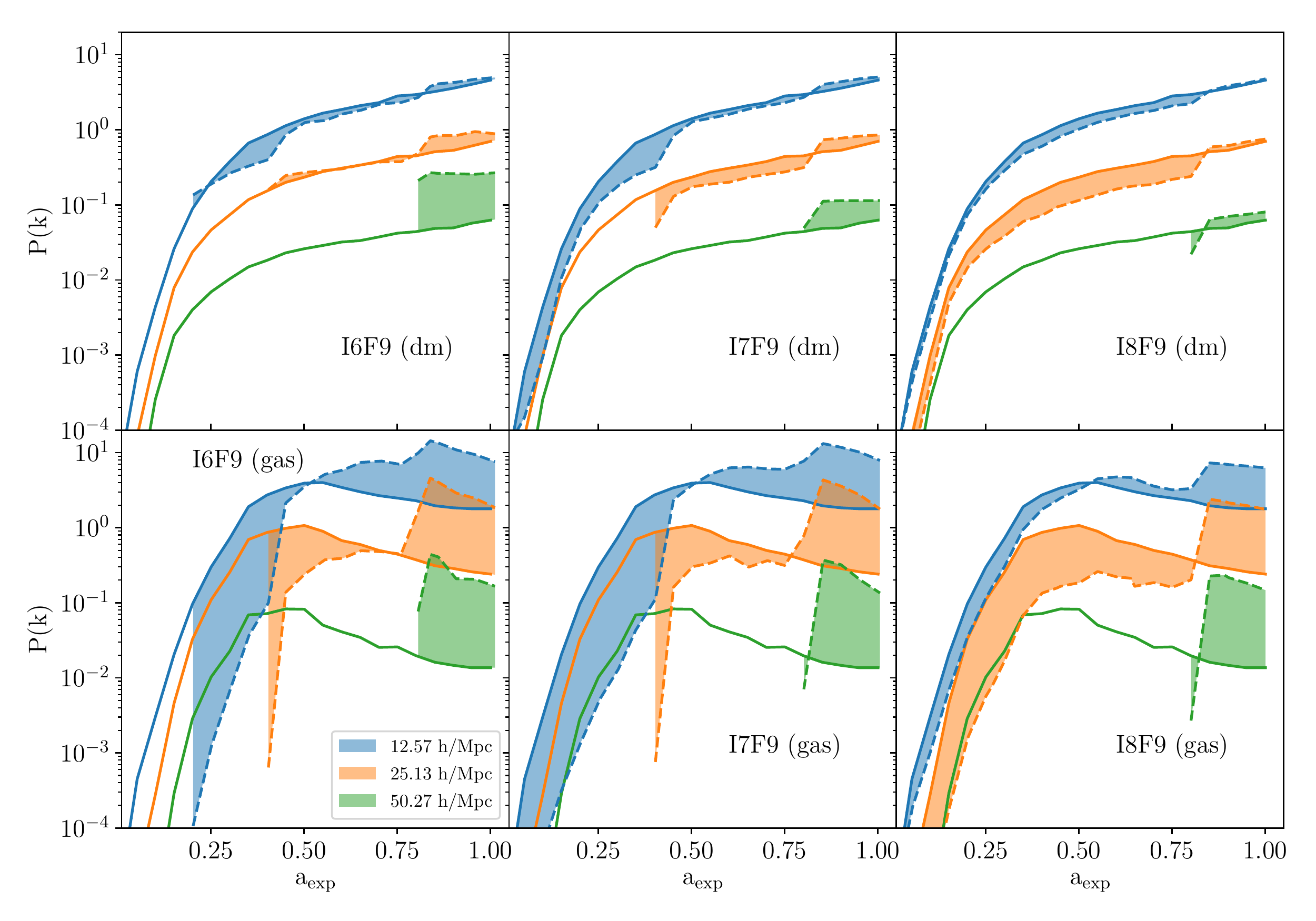} 
\caption{The evolution of the power spectral amplitude measured at three fixed scales of $k$ against the expansion factor. The top and bottom rows show the evolution for dark matter and gas, respectively. From the left column the simulations I6F9, I7F9 and I8F9 are shown. In each panel the dashed line shows the power spectrum for the associated AMR simulation (e.g. I6F9, I7F9 and I8F9) while the solid line show the evolution of the fixed grid I9F9 simulation. For clarity we shade the region of difference between IxF9 and I9F9. The values of $k$ are coloured to match the points in Fig. \ref{Fig:Figure1b} which show the associated $k$ value. The dashed lines are plotted only when the simulation has reached a sufficient grid level that there is information on that scale. If the force resolution is lower than the resolution of the fixed grid used to calculate the power spectrum, then no line is shown. In the fixed grid case the simulation always has this resolution. The title of each column describes the line colour of the represented simulation in Figs. \ref{Fig:Figure1} and \ref{Fig:Figure1b}. }
\label{Fig:Figure2}
\end{figure*}

Figure \ref{Fig:Figure2} takes four values of $k$, marked by the three points in the lower panel of Fig. \ref{Fig:Figure1b} and traces their evolution through time. The evolution of the power spectrum over time is smooth only for the fixed grid (dashed line) in both the dark matter and the gas. The AMR simulations (solid lines) show evolution punctuated by large changes in the gas power spectrum (and lesser ones in the dark matter), and each simulation shows these jumps at predefined intervals. These intervals are fractions of the expansion factor, $a_{\mathrm{exp}}$. 

The smaller-scale power evolves in step with the number of grids because of the restriction in the maximum refinement level discussed above \citep[e.g.][]{Dubois2012}. Once a new level is allowed, the initial power on that level is too low because the initial density fluctuations on this scale are due only to the initial interpolation (in our case this is a simple step function). However, the result of the simulation is not strongly influenced by any well behaved interpolation \citep[see][]{Rabold2017}. Once that level is allowed, further fluctuations are able to build up due to various gravitational and hydrodynamical processes. The last refinement level is allowed after $a_{\mathrm{exp}}\sim$0.8 for each of the AMR runs. For I8F9 this is the only refinement release, because there is only one allowed refinement level,which is held back until this epoch. Before this epoch, the simulation functioned as a I8F8 fixed grid. Importantly, only the last refinement level release is responsible for the high $\delta^{\mathrm{AMR}}_{\mathrm{ref}}$ at $z$=0 as the last jump in power makes all the AMR power spectra have similar distributions, (essentially erasing the previous distribution) and sets the shape of the $z$=0 power spectrum in each of the AMR runs.  

The release of the highest refinement level at $a_{\mathrm{exp}}\sim$0.8 disturbs the system and builds up spurious small-scale power, of approximately the same magnitude in each AMR run. Furthermore, the shock to the system of releasing that last refinement level affects the power on scales larger than the newly allowed $k$ values. This is because \textsc{RAMSES} does not allow jumps of more than one level from one neighbouring cell to another, meaning that intermediary grids must be formed. Additional grids at levels 8, 7 and 6 (for example) must be produced in order for some of the level 9 grids to be generated. This occurs once the expansion factor passes a certain threshold, e.g. between one time step and another, so the system {\it suddenly} changes, resulting in a sharp change to the number of degrees of freedom. 

\begin{figure*}
\centering
\includegraphics[scale=.46]{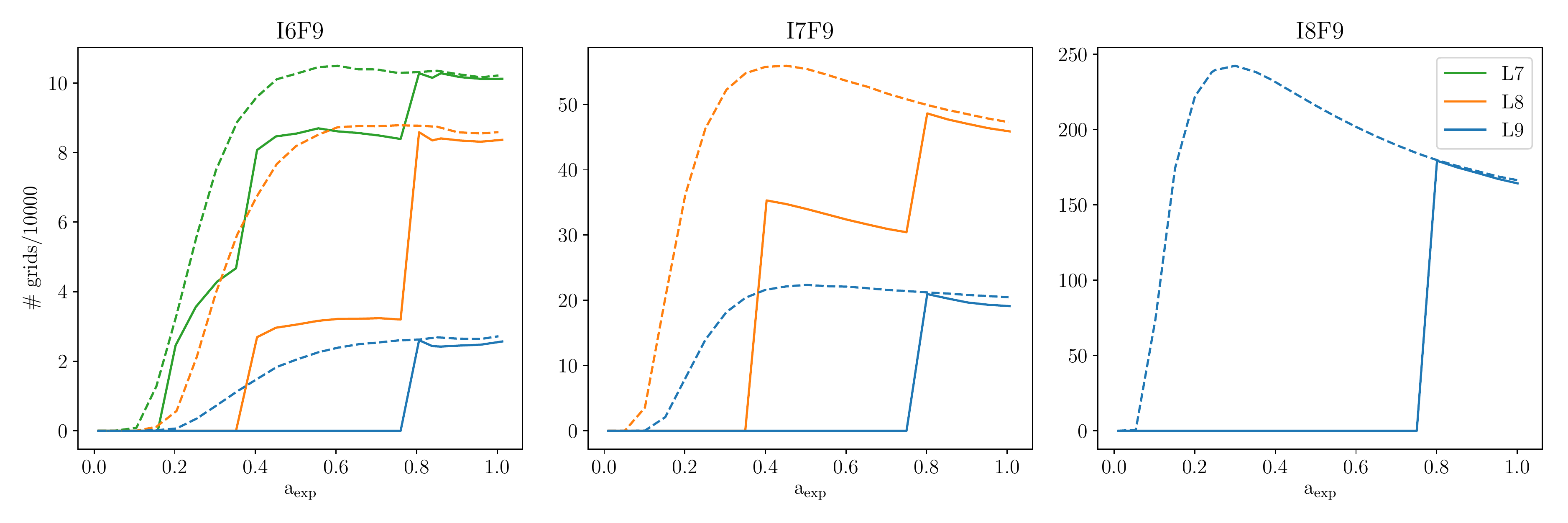} 
\caption{The inherent differences in the evolution of grid populations at each refinement level with and without star formation. In each panel we show the evolution of the number of generated grids with the star formation mode ({\it solid lines}) contrasting with the runs without star formation ({\it dashed lines}; see text for details). Note that before $a=0.8$ no grid refinements at this level are allowed in the star-forming case, and no grids in level 8 are allowed before $a=0.4$ in any of the star forming simulations. At the epoch when the restriction on the refinement is lifted, there is a noticeable jump in the number of grids even in lower levels. This is required to prevent grid cell size differences of more than one level in high density regions. The non-star-forming case does not show the same hold-back. The panels show which of the different IxF9 simulations applies in each panel. Only levels which are produced by refinement are shown. }
\label{Fig:levels}
\end{figure*}

This same effect is not as important in the dark matter because, although the force resolution increases, the number (and mass) of particles in the system remains constant. In the hydrodynamics, however, many more cells are produced on multiple levels other than the one released (solid lines, Fig. \ref{Fig:levels}). Grids can be seen to decrease for I7F9 and I8F9 because isolated regions of high refinement (e.g. halos) merge over time due to hierarchical merging and void formation.

\subsubsection{Simulations without hold-back}
\begin{figure}
\centering
\includegraphics[scale=.45]{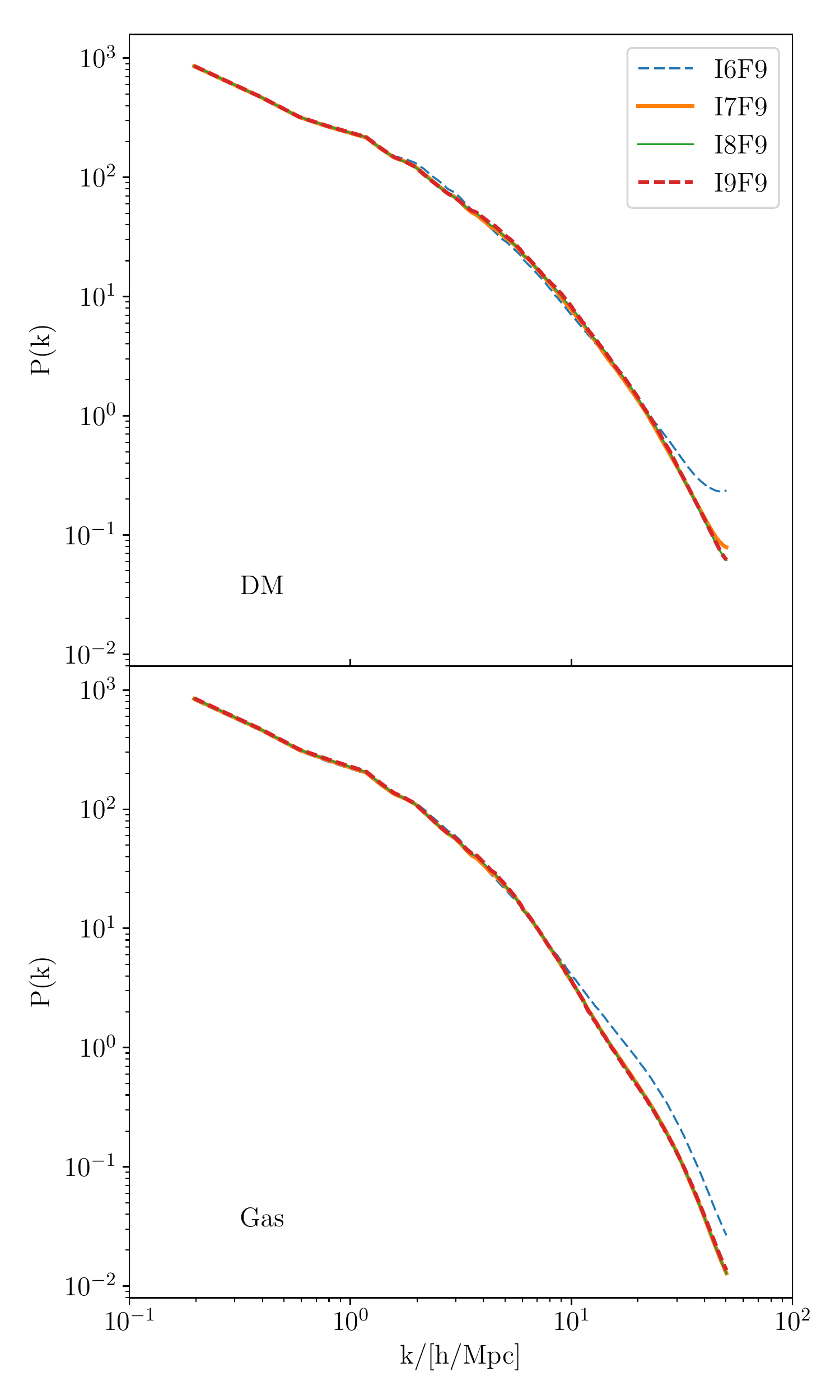} 
\caption{Same as Fig. \ref{Fig:Figure1}, but for the no-star formation \& no-hold back case.  }
\label{Fig:Figure4}
\end{figure}

\textsc{RAMSES} holds back levels to ensure constant physical resolution. However, this is only true when \textsc{RAMSES} is asked to form stars.  Without the effectively constant physical resolution at early times there will be insufficient mass in a cell at high redshift to correctly form stars. When \textsc{RAMSES} is in pure hydrodynamics mode (i.e. star formation is forbidden by setting $\epsilon_{star}$=0) this restriction is not implemented. In this case, the system does not experience the same sudden, and global, change in the number of grids. If we carry out the same I6F9, I7F9, I8F9 and I9F9 simulations as above but explicitly turn off star formation (and grid hold-back), the number of grids evolves smoothly (dashed lines, Fig. \ref{Fig:levels}) in each simulations even if the exact evolution of the number of grids changes with resolution. For example, there are far more grids overall in level 9 in the I8F9 simulations than in the lower resolution runs because more particles are present to trigger refinements. Furthermore, the number of grids at level 9 in the I8F9 run peaks at a$_{\mathrm{exp}}=0.25$ and then falls at later times.
Ultimately, however, the two modes reach approximately the same number of grids by $z$=0 but the additional small scale power in the gas seen in Fig. \ref{Fig:Figure1} is not apparent (lower panel Fig. \ref{Fig:Figure4}).

\begin{figure}
\centering
\includegraphics[scale=.45]{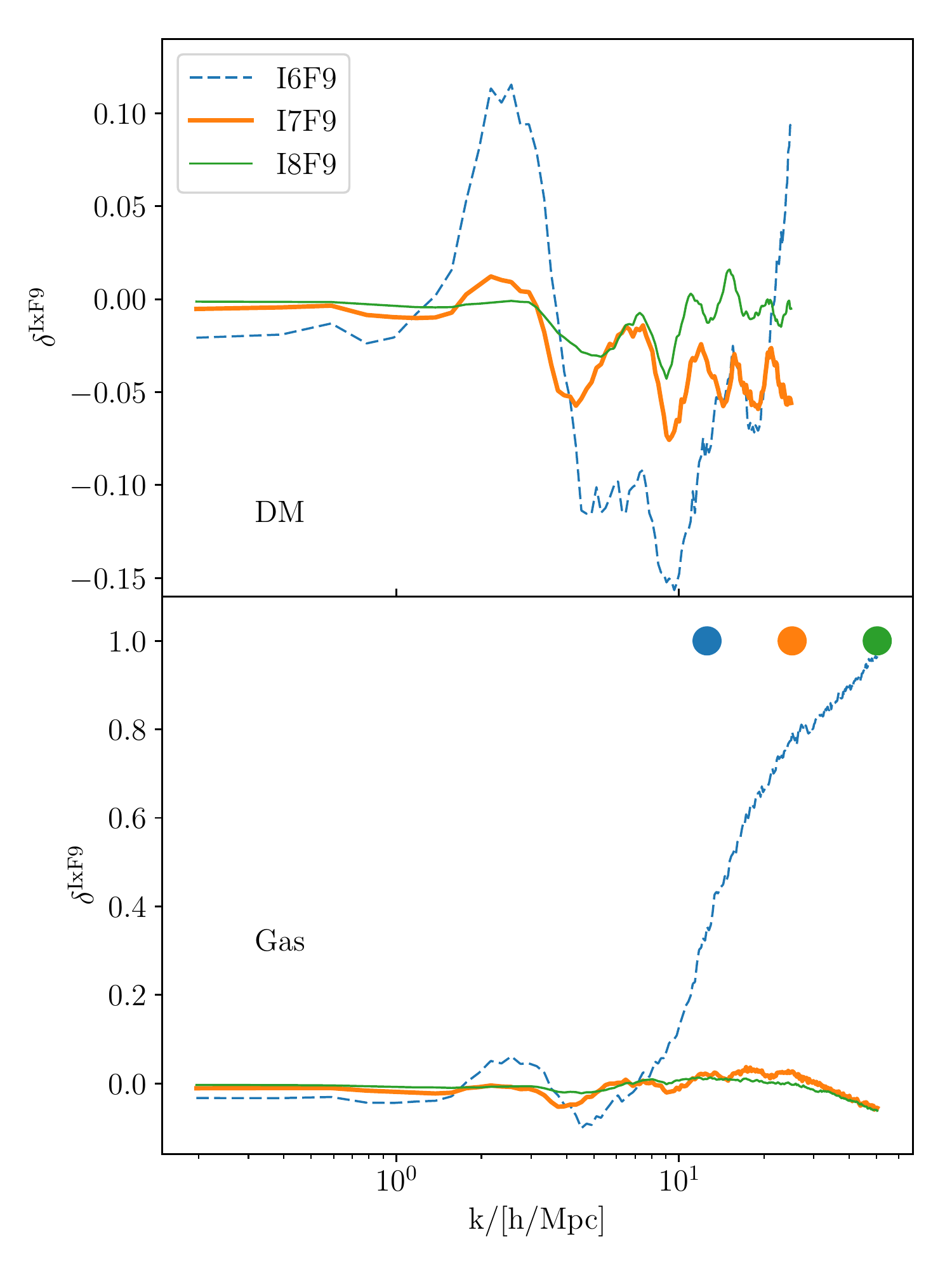} 
\caption{Same as Fig. \ref{Fig:Figure1b}, but for the no-star formation \& no-hold back case. Everything else is the same. The top panel shows the difference between IxF9 and I9F9 for the dark matter, and the middle panel shows the same for the gas.}
\label{Fig:Figure4b}
\end{figure}

Figure \ref{Fig:Figure4} shows the power spectra of the four simulations for the gas and dark matter when the star formation is turned off and grids are not held back. There are substantially more grids, and so degrees of freedom in the gas, at early times when no star formation takes place. In this case the different IC resolution runs converge with increasing resolution as initially predicted. This is clearly demonstrated in Fig. \ref{Fig:Figure4b}, which shows a smaller $\delta^{\mathrm{AMR}}$ for the dark matter power spectra and a considerably smaller (an order of magnitude compared to the star forming case) $\delta^{\mathrm{AMR}}$ for the gas. { This difference is not due to star formation physics as at IxF9 resolution, and star formation would be expected to reduce the small scale power, not enhance it.}

This has important consequences for the choice of refinement strategy, and demonstrates that large global refinements cause considerable artefacts in the system as a result of the approach to refinement. When the grids are free to form according to the density distribution alone, the evolution is as expected, and the increasing resolution of the initial conditions causes a smooth convergence towards the reference model. 

\begin{figure*}
\centering
\includegraphics[scale=.6]{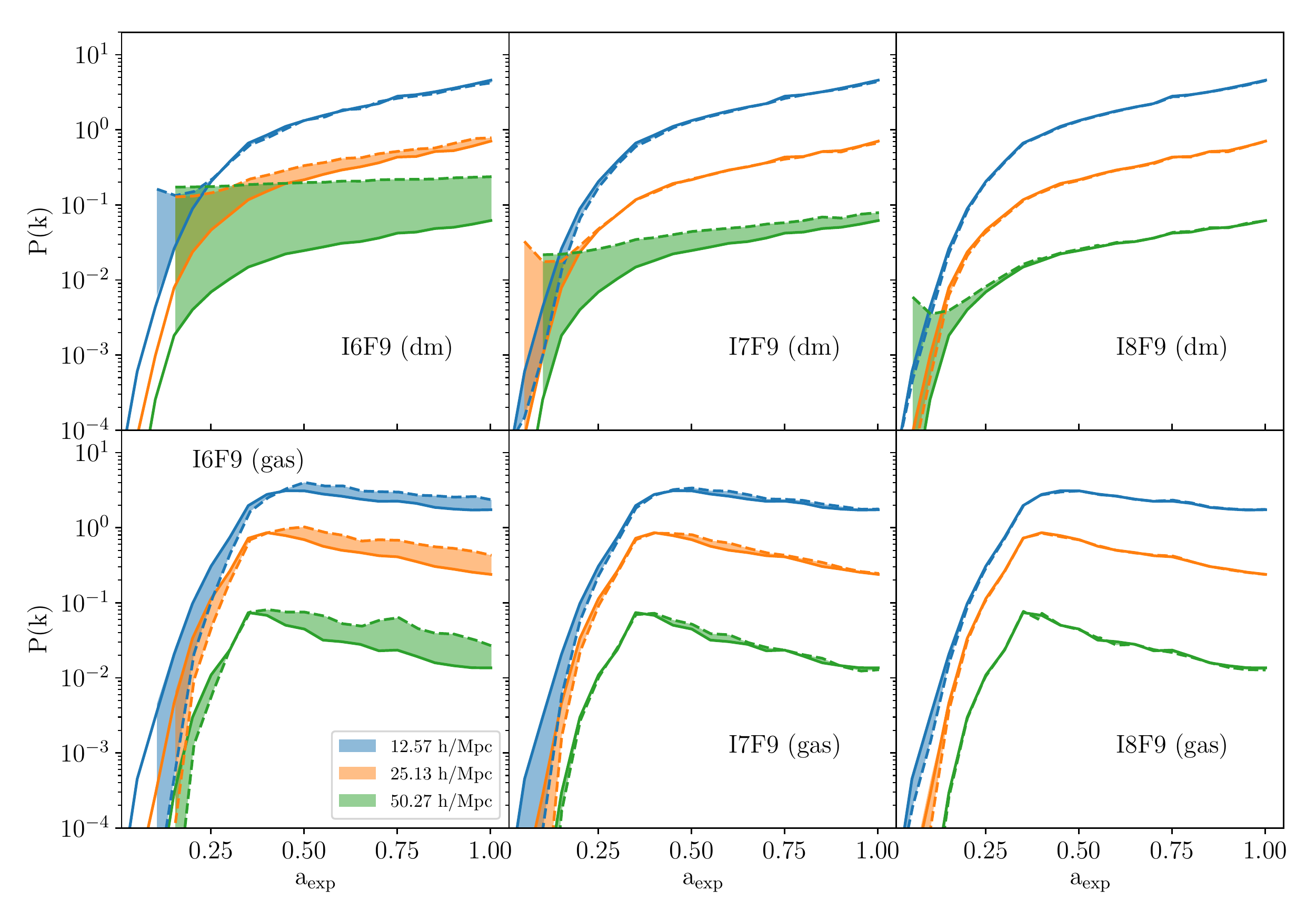} 
\caption{Same as Fig. 4, but for the no-star formation, no-hold back case.}
\label{Fig:evolNS}
\end{figure*}

The evolution of the non-star forming material is also of interest. For the dark matter, the largest $\delta^{\mathrm{AMR}}$ at $z$=0 occurs mainly close to the Nyquist frequency (shown in Fig. \ref{Fig:Figure4b}). However, for I6F9 and I7F9 the small-scale power in the dark matter is much higher compared to the reference model at early times and barely evolves over time (Fig. \ref{Fig:evolNS}, top row). This is because even though grids are allowed to form on small scales, the dark matter particles have not sufficiently clumped at early times to produce power on the smallest scales measured, and the apparent power on these scales is numerical rather than physical. Once the particles have become sufficiently clumped that the power at high $k$ is meaningful, the AMR and reference runs grow similarly, as shown in the I8F9 run.  

The gas power spectrum shows a sharply different evolution. As there is no discretion into particles for the gas, there is meaningful information as soon as the mesh has refined to the level corresponding to that $k$ value. The dark matter power reaches a maximum value at $z$=0, while the gas peaks at $a_{\mathrm{exp}}\sim0.3$ and then falls again. This is true at all $k$ and for all runs. There is less power at high $k$ in the AMR runs compared to the reference with falling $\delta^{\mathrm{AMR}}$ as the resolution increases as shown in Figs. \ref{Fig:Figure4} and \ref{Fig:Figure4b}. But this is true at all times throughout the run. 

Where the grids are not held back,  and the adaptive mesh is free to refine according to the pseudo-Lagrangian method alone, the different evolutions of gas are much closer to the reference model.

However, in the star-forming case, the gas showed the largest offset from the reference model without star formation, the dark matter shows the strongest offset, and this offset is strongest at early times \citep[c.f.][]{OShea2005}.

\subsection{Simulation force resolution}

One issue with the above analysis is that there are few available refinement levels for the AMR to use. This limited simulation resolution was chosen so that a fixed grid could be used as a reference model. However, many \textsc{RAMSES} simulations use around 7 refinement levels \citep[e.g.][]{Dubois2016}. Thus, in order to test the effect of the IC resolution on the outcome of a more realistic \textsc{RAMSES} simulation we allow refinements up to level 15. It is clearly impossible to generate a fixed grid reference model down to these resolutions, and would involve 32768$^3$ grids, which is beyond the reach of current computing facilities. We continue to calculate the power spectrum on the 512$^3$ grid used in the previous section. The simulation resolution is also a vital factor in determining the endpoint power spectrum of the simulation as discussed by many authors \citep[e.g.][]{Teyssier2002}. In order to explore the impact of the simulation resolution we ran a new set of simulations I7F10, I7F11 and I7F12 and I7F15. The I7F15 simulation has a force resolution of $\sim$1 $h^{-1}$kpc, while I7F12 has a force resolution of $\sim$8 $h^{-1}$kpc.  The $F15$ resolution is comparable to the gravitational softening used in Illustris-2 for the gas particles, and Illustris-1 for dark matter. The two, however, are not entirely equivalent because the minimum gas cell size used is an order of magnitude smaller for the Illustris gas smoothing scale than the softening scale, while in \textsc{RAMSES} the two are equivalent.

\begin{figure}
    \centering
    
     \begin{tabular}{cc}

     \includegraphics[scale=.6]{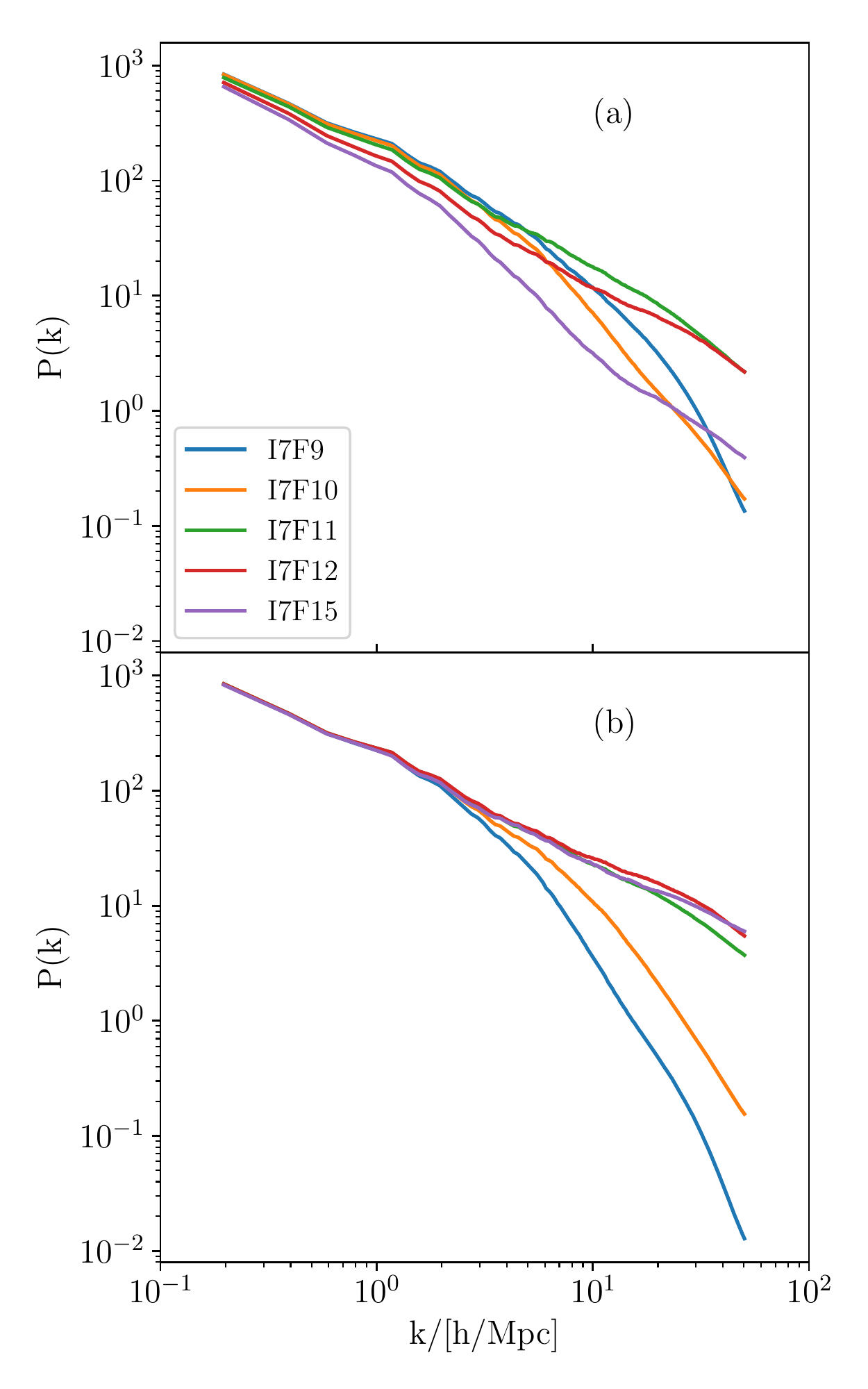} \\
     \includegraphics[scale=.5]{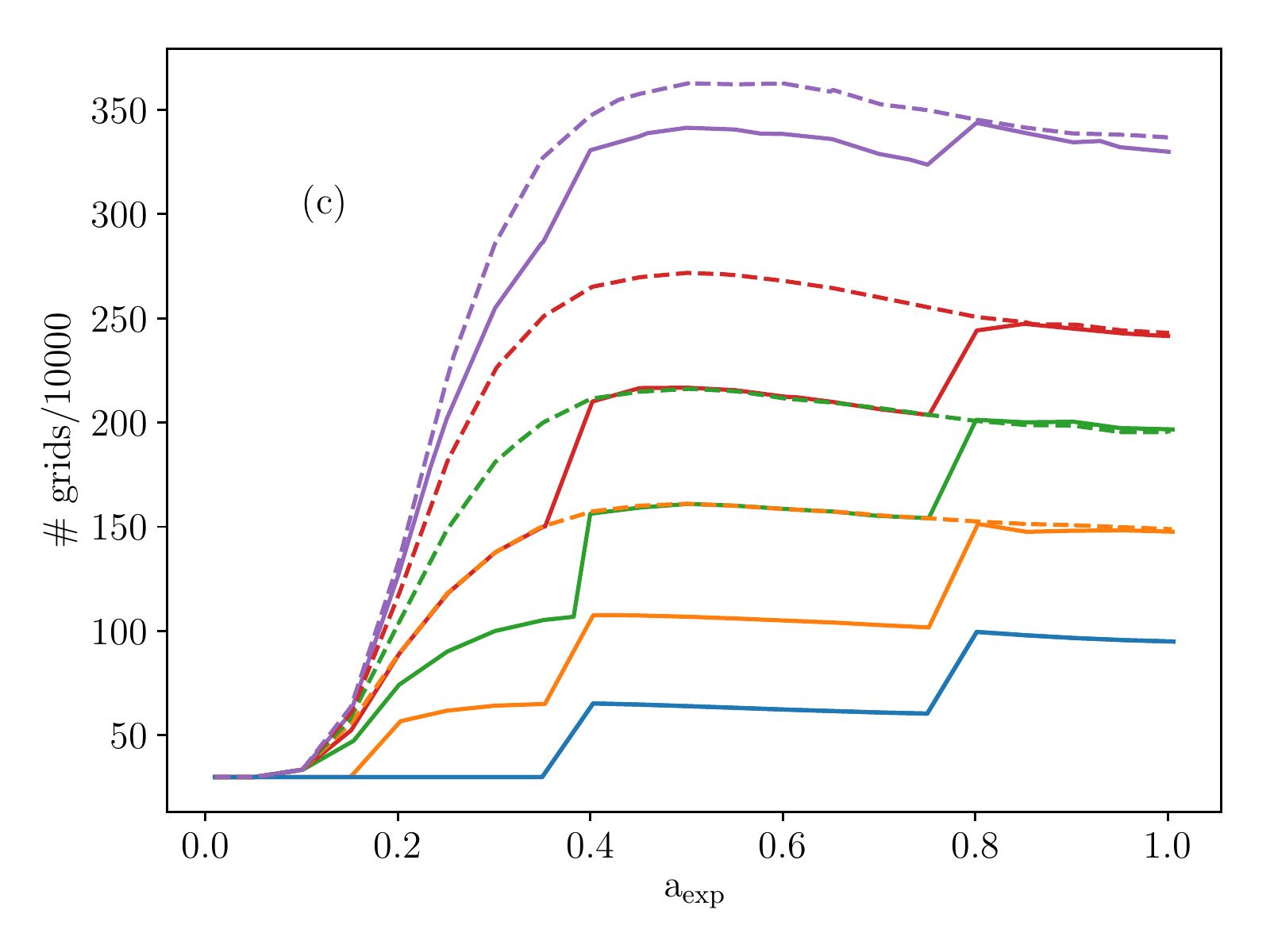} \\ 
     (c) \\         
     \end{tabular} 
     \caption{Panel (a): The $z$=0 power spectra for I7F9, I7F10, I7F11, I7F12, I7F15 with star formation. Panel (b): the same as panel (a) but without star formation.  Panel(c): The number of grids against the expansion factor for both the star-forming (solid lines) and non-star-forming runs (dashed lines).}
     \label{Fig:SimresPkz0a}    
\end{figure}

\begin{figure}
\centering
\includegraphics[scale=.6]{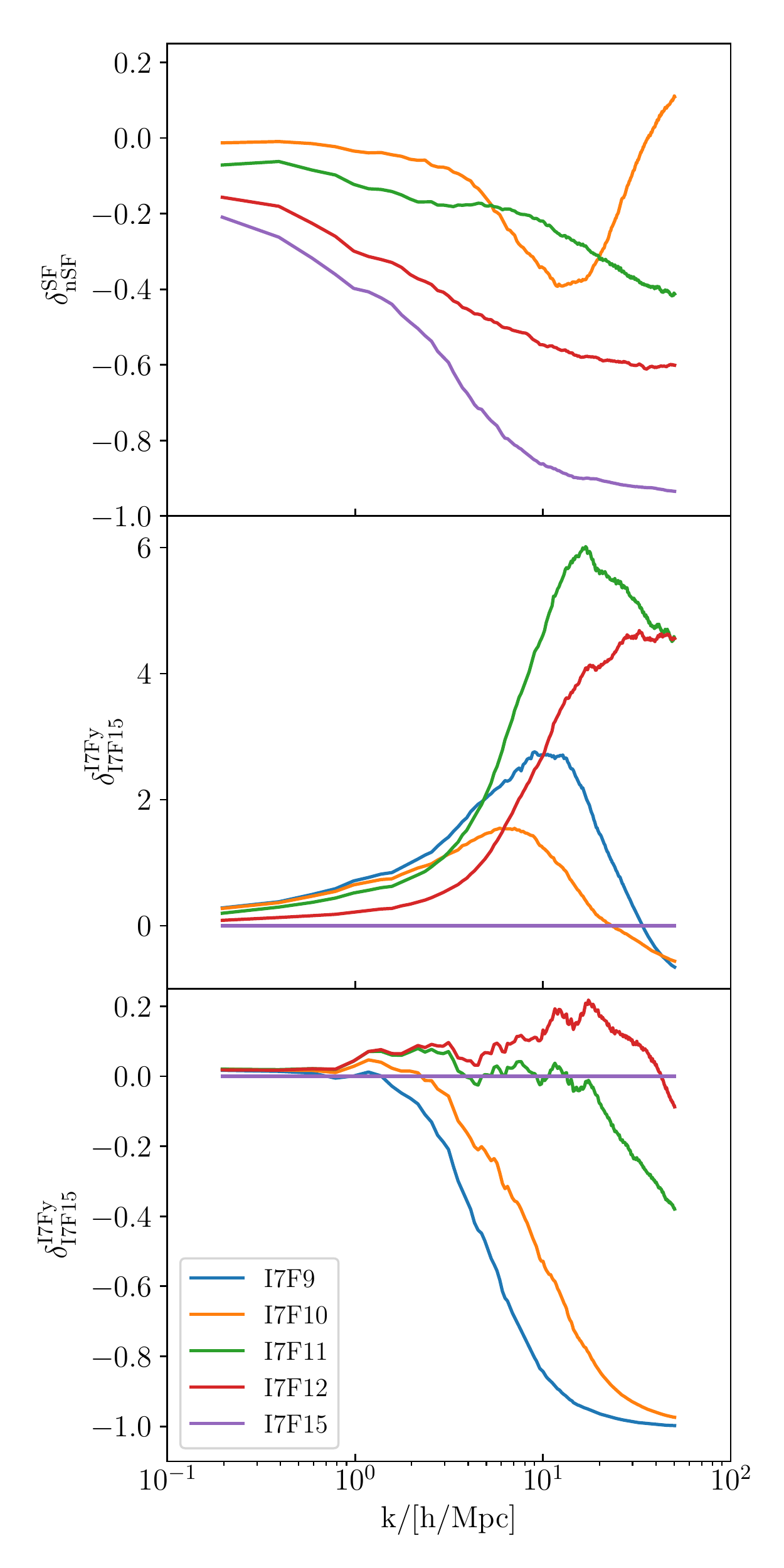} 
\caption{Power spectral contrast of gas among I7Fy for star forming and non-star forming runs. Top panel: Difference between the power spectra for simulations with and without star formation at the same resolution. Middle panel: Comparison between power spectra for the different star forming simulations relative to I7F15 (star forming), Bottom panel: Same as the middle panel but for the non-star forming runs. I7F9 is missing from the upper panel because it is difficult to see the evolution of $\delta$ for the other runs if it is included due to the large offset at high $k$.}
\label{Fig:evolNSdelta}
\end{figure}

For the gas, the difference between the cased with and without grid hold-back  is very large, (compare panels (a) and (b) in Fig. \ref{Fig:SimresPkz0a} and the top panel of Fig. \ref{Fig:evolNSdelta}) while the difference is less evident in the dark matter. This is unsurprising as the dark matter is not directly affected by feedback (though it will be influenced by the redistribution of baryons). At this point we will concentrate on the gas as there is an extensive literature on the behaviour of the dark matter \citep[e.g.][]{LHuillier2014}. 

The insight that these higher simulation resolution runs have more power on small scales relative to the reference simulation (I9F9) is unsurprising. This is because the higher simulation resolution is expected to naturally produce more small-scale power over the evolution \citep[e.g. Fig 11,][]{Teyssier2002}. If we assume that the highest resolution simulation (I7F15) is the most accurate simulation of the Universe, we can compare each simulation to this new reference simulation.  See section \ref{Subsec:ICres} for a more detailed discussion of the limits of the final spatial resolution for a given set of ICs.

The lowest resolution (I7F9, I7F10) simulations have less power on small scales than the highest resolution case, and this is exactly as expected; the more degrees of freedom in the system, the more small scale power we expect to build up. As the resolution increases, there is initially an increase in the power at high $k$ (from $\sim$2 to $\sim$6). However, between I7F12 and I7F15 the power falls again (from $\sim6$ to $\sim5$), such that I7F15 has the lowest power of all the runs. This is illustrated in the middle panel of Fig. \ref{Fig:evolNSdelta}, which shows a growing $\delta^{\mathrm{I7Fy}}_{\mathrm{ref}}$ ($\delta^{\mathrm{I7Fy}}$, henceforth) from I7F9 to I7F11 and then a decrease from I7F11 to I7F12. Although at the highest $k$, I7F9 and I7F10 have less power than I7F15 because, for most values of $k$ there is more power.

The gas without  hold back (or star formation) clearly shows a much stronger difference from the runs with  grid hold-back  as we increase the resolution (bottom panel of Fig. \ref{Fig:evolNSdelta}). This is because there has been more time when the highest resolution levels were available, effectively meaning that the co-moving resolution was higher at early times, allowing extensive cooling (overcooling) to take place. There is on average less power in the star forming case than in the no-hold-back case at the same resolution (top panel, Fig. \ref{Fig:evolNSdelta}). 

There is a strong influence of resolution on the power spectrum without hold-back, but it is markedly different to the with-hold-back case. As we increase the resolution, there is more power on small scales in each case, and as the resolution increases the difference between the lower resolution runs and the reference decreases and converges. 

For I7F9 and I7F10 the maximum $\delta$ is approximately -1.0 at $k=1.8$ $h$/Mpc, while for I7F12 it is -0.15 (peaking at +0.1 at intermediate $k$). This difference is a lot smaller than in the hold-back case. Furthermore, the trend in $\delta^{\mathrm{I7Fy}}$ does not turn over. This is because the hold-back case produces stars more efficiently at higher force resolution and so additional physics affects the outcome while at lower resolution the effect of hold-back is more important (lower panel Fig. \ref{Fig:SimresPkz0a}).

\subsection{IC resolution at higher simulation resolution}
\label{Subsec:ICres}

The maximum simulation resolution has a strong effect on the predictions of the simulation in both the dark matter and gas, while not all these trends are obvious. It is vital to extend the work in \S \ref{Sec:Default} to a resolution more similar to most current hydrodynamical simulations \citep[e.g.][]{Vogelsberger2014}. Thus, we carry out the same procedure for a simulation resolution of 1 $h^{-1}$kpc, keeping the same box size as in the previous sections. Due to the relatively high computational cost we omit the I9F15 simulation, leaving I8F15 as the simulation with the highest IC resolution. We carry out the simulation with and without  hold-back as before. The lack of star formation in the  non-hold-back case enables the gas to generate more power on smaller scales, because there is nothing to resist cooling where no stars are formed. Indeed, this is a manifestation of the `overcooling problem' \citep{Rasera2006}, which is overcome via strong feedback (see the preceding section and Fig. \ref{Fig:SimresPkz0}).

\begin{figure}
    \centering
    
     \begin{tabular}{c}
     \includegraphics[scale=.4,trim={.5cm .0cm .5cm 0cm},clip]{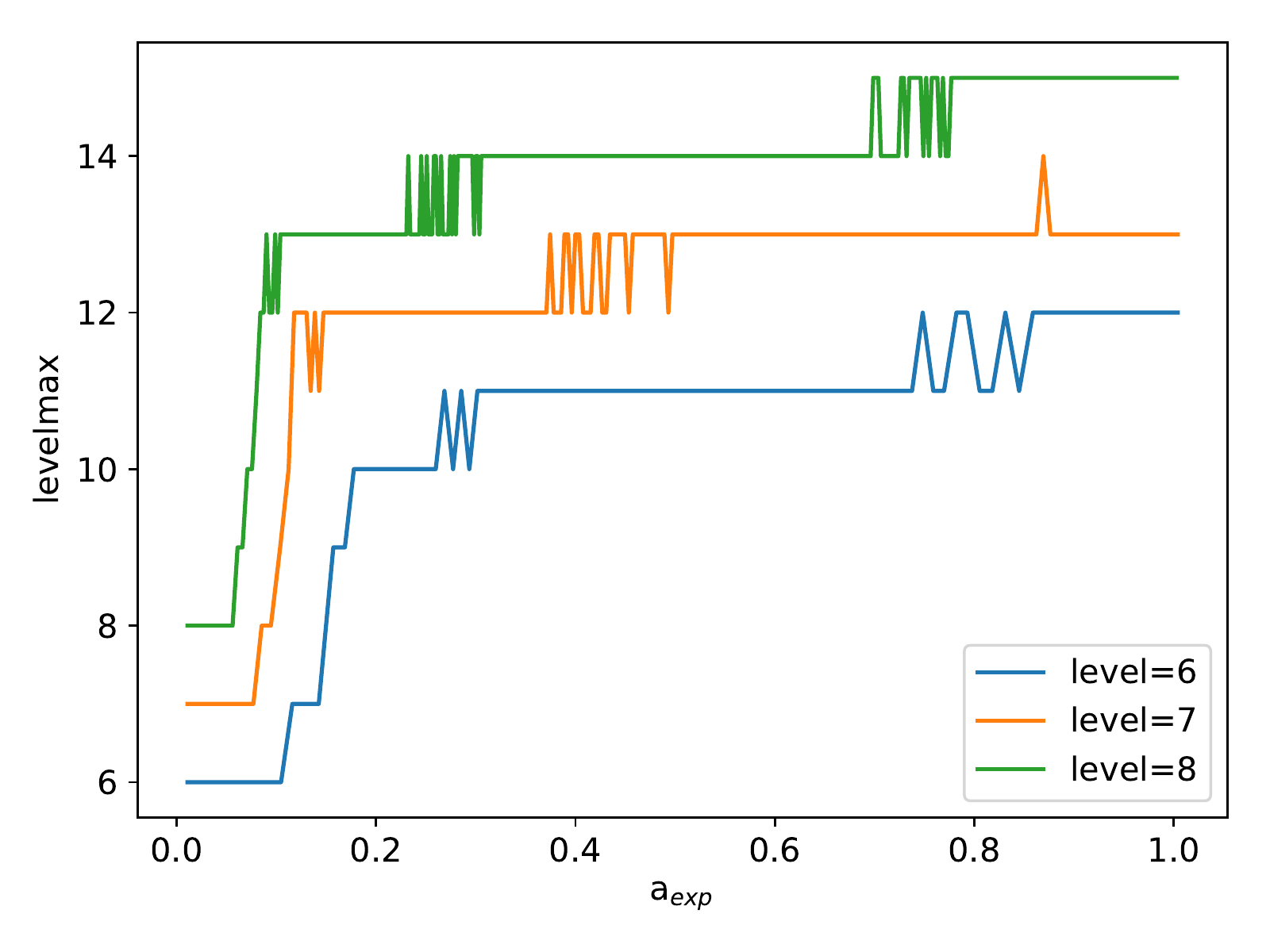}\\    
     \end{tabular} 
     \caption{The evolution of the maximum effective refinement level of the dark matter only simulation for I6, I7 and I8.}
     \label{Fig:DMonlylev}    
\end{figure}

\begin{figure}
    \centering
    
     \begin{tabular}{c}
     \includegraphics[scale=.6,trim={.5cm .5cm .5cm 0cm},clip]{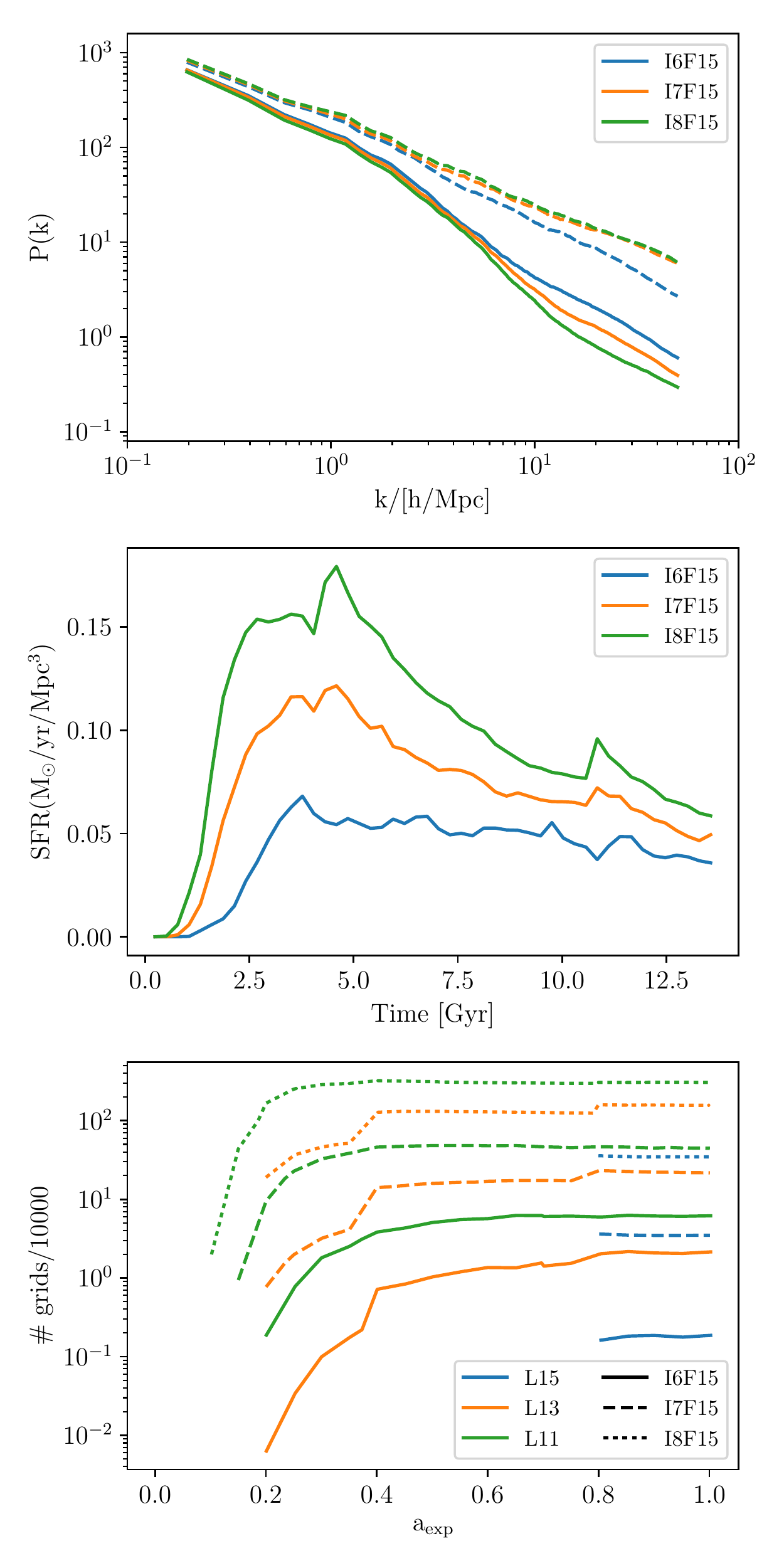}\\    
     \end{tabular} 
     \caption{Top panel: The power spectrum of gas in the simulations IxF15, where x is 6,7 or 8. The solid lines include star formation, and hold back the grids to ensure a similar physical resolution. The dashed line has no star formation and no hold back on the grids. The initial conditions are the same as shown in Figure 1. Note that the non-star forming case converges to a given profile as the IC resolution increases. The star-forming case does not. The higher resolution ICs produce more power on smaller scales in the non-star forming runs but the opposite behaviour is seen in the star-forming case. Middle panel: The star formation rate of I6, I7 and I8. Bottom panel: The number of grids in several levels in the I6, I7 and I8 runs. The higher resolution ICs have more particles, therefore more grids can be refined and the density more accurately probed. The number of levels shown are distinguished by line colour and the IC resolution by the line style, as shown in the legend.}
     \label{Fig:SimresPkz0}    
\end{figure}

 We have increased the maximum force resolution to F15. This is not ideal for the lower resolution initial conditions (e.g. I6) because the resulting dynamic range is very high (a factor 512). If we carry out a dark matter-only simulation we can see that I6, I7 and I8 naturally reach between six and seven additional levels of refinement. The fact that the I6 simulation is able to achieve level 15 in full hydrodynamical mode means that these extra two levels are only due to the effect of the hydrodynamics, and not the natural distribution of dark matter clustering. The higher resolution can result in numerical effects caused by particle-particle scattering and collisions. The levelling over time of the dark matter only simulation is shown in Fig. \ref{Fig:DMonlylev}.  Thus, the I6F15 simulation in particular may show unphysical properties.\footnote{ \textsc{RAMSES} is capable of allowing different grid levels for dark matter and hydrodynamics, using the \textsc{cic\_max} parameter. This can be set to be lower than the \textsc{levelmax} parameter which is used to define the maximum spatial resolution.} In this paper, however, we intend to show the impact of the resolution on the outcome of simulations, so although the reader is invited to remain aware of this caveat, we will carry out simulations to F15 so that we have the maximum possible dynamic range. We will discuss carrying out simulations to F12 and F13 respectively in the appendix.

The non-hold-back simulations demonstrate strong convergence with the I6F15 simulation having less power ($\delta^{\mathrm{I6F15}}_{\mathrm{I8F15}}=-0.56$) on small scales than the I7F15 and I8F15 simulations, and the I7F15 has slightly less power than the I8F15 simulations ($\delta^{\mathrm{I7F15}}_{\mathrm{I8F15}}=-0.11$). Thus, the highest resolution simulation is able to form smaller structures with a greater density contrast. This is also clear for IxF12 and IxF13 runs shown in the appendix.

Conversely, where star formation occurs, there is a considerable reduction in the small-scale power. This is because the stellar feedback acts on the ISM to prevent cooling and collapse which occurs in the non-star-forming case. The difference in I8F15 with and without star formation is $\delta^{\mathrm{SF}}_{\mathrm{nSF}}=20$ (SF is star forming (held back) and nSF is non-star forming (non-held-back)), which is considerable. However, we see the opposite trend compared to the simulation resolution result in the  non-held-back runs because the IC resolution and smaller scale power are anti-correlated. We can see that the crossover occurs between IxF13 and IxF15, because the order of the power spectra at high $k$ swaps between these two spatial resolutions for the  held-back case. This suggests that the effect of star formation is dominant over the IC resolution. But it should be remembered that we don't have the simulation with the full initial resolution and fixed grid that can be used as the reference. 

The critical difference among the  grid hold-back runs appears to be the total mass of stars formed, shown in the middle panel of Fig. \ref{Fig:SimresPkz0}  and Figs. \ref{Fig:Restest12} and \ref{Fig:Restest13}. As the amount of energy released into the gas from stars is a function of the mass of stars formed, the higher star formation rate runs will produce less small scale power  on intermediate scales. However, on smaller scales higher star formation may result in clumpier gas and so higher small scale power. We do not measure the power spectrum to these scales, however. The middle panel of Fig. \ref{Fig:SimresPkz0} demonstrates that the I8F15 simulation has a higher star formation rate than I7F15 or I6F15  (or any of the runs in the appendix). This higher resolution simulation, therefore, removes more gas from the ISM (in order to form stars), and heats the gas via feedback. This means it is more difficult to form the next generation of stars. Furthermore, although changing the star formation efficiency, or density threshold, changes the shape of the star formation history, the overall amount of stellar mass formed remains smaller in the lower resolution runs compared to the higher resolution runs for a range parameters (e.g. the star formation efficiency ranges from 0.02-0.2 but the total mass of stars does not reach I7F15 from I6F15). 

One of the reasons for this effect is that the simulations with the higher resolution ICs have more particles. As the chosen refinement criteria are the 'Quasi-Lagrangian' type, where refinement is controlled by the number of particles in a grid cell, the more particles in the system the more readily the simulation can refine. As the resolution decreases, the (dark) matter density field is traced using fewer and fewer particles, which means that there is much less chance of getting eight of those particles in a cell. This is reflected in the bottom panel of Fig. \ref{Fig:SimresPkz0}, which shows the number of grids in refinement level for each of the simulations. I8F15 has far more than I6F15, for example, because of this. Thus, more grids in the I8F15 run have sufficient density to form stars compared to I6F15, and, therefore, the star formation rate is higher. 

\begin{figure*}
    \centering
    
     \begin{tabular}{cccc}
     \includegraphics[scale=.4]{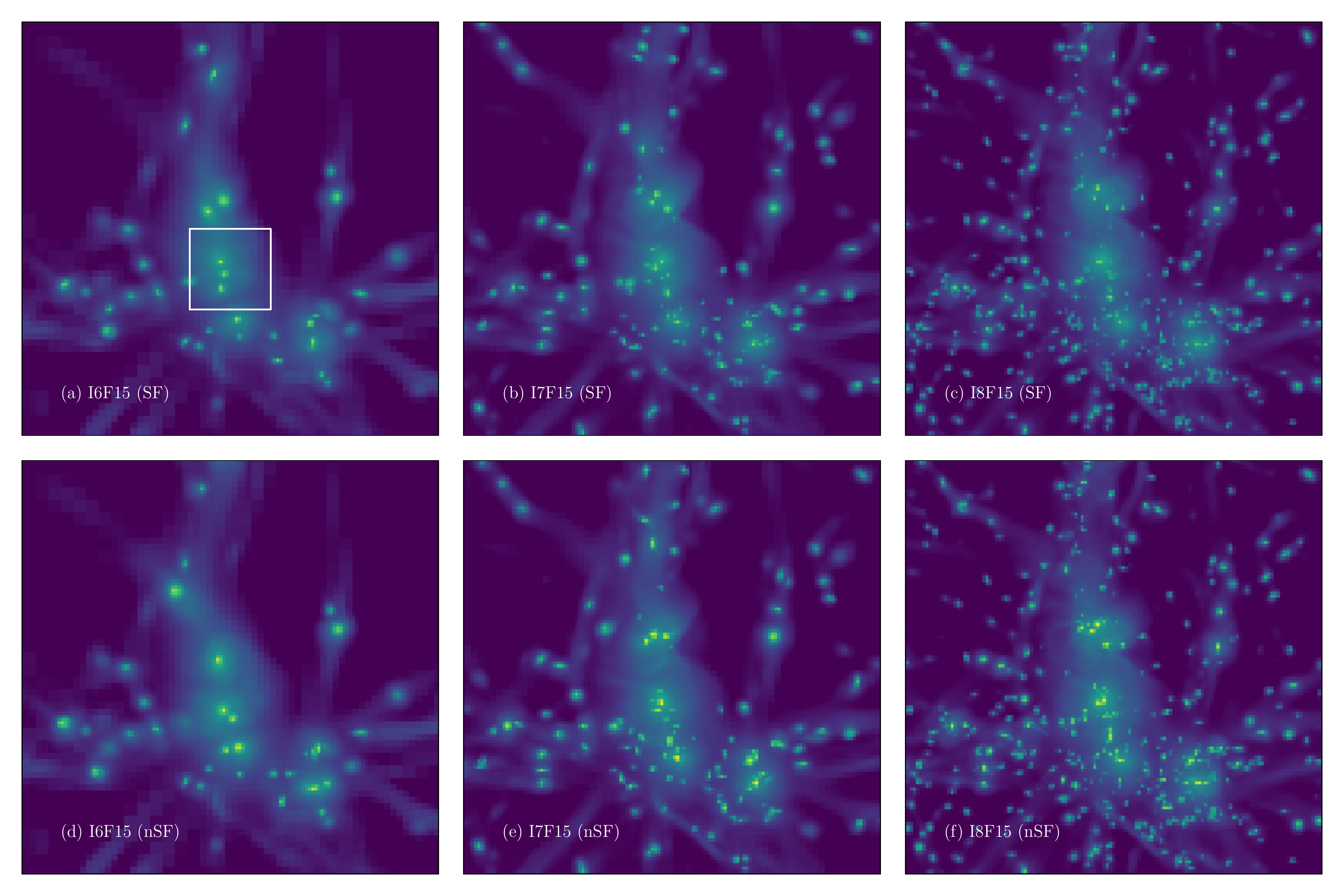}\\ 
     \hline 
     \includegraphics[scale=.4]{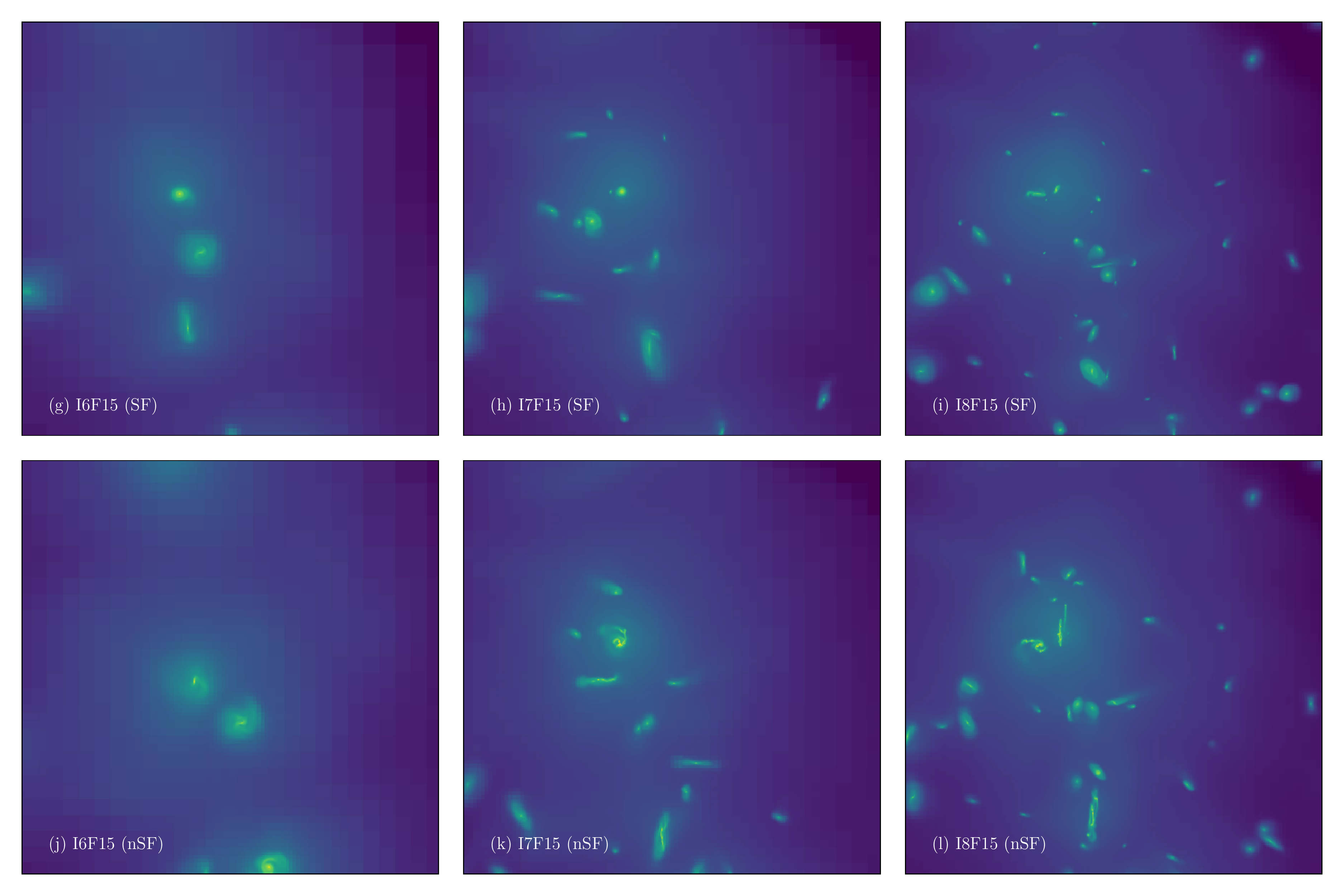}\\                
     \end{tabular} 
     \caption{Effect of increasing the IC resolution on the outcome of the simulation. The gas density distribution of the volume near the most massive halo (with a mass of 6.4$\times 10 ^{13}$h$^{-1}$M$_\odot$) which is 8 $h^{-1}$Mpc (1.6 $h^{-1}$Mpc) wide in the top (bottom) two rows in projection. Panels a, b, c in the upper two rows show the runs with star formation, while panels d, e and f show the distribution for the no star formation, free grid run. From left to right the I6F15, I7F15 and I8F15 runs are shown. In panels a-f the density is taken only down to F9 so the same information is available in these plots as is used to produce the power spectra in Fig. 10. In panels g-l we shows a zoomed in region 1.6 $h^{-1}$Mpc down to level 15 (this region is marked as the white square in panel a). 
     }
     \label{Fig:image}    
\end{figure*} 

Figure \ref{Fig:image}  zooms in on a subregion of the simulation volume for the different runs for which a power spectrum is presented in Fig. \ref{Fig:SimresPkz0}, and visually displays the gas distribution. It is instantly clear that even at fixed force resolution, the resolution of the initial conditions has a striking impact on the final gas distribution. For I6F15 there are far fewer small scale clumps than in the higher resolution simulations regardless of whether star formation takes place or not. It is therefore vital to supply both the initial and final resolutions of simulations in order to make valid comparisons between different simulations. With the absence of small scale structure in lower resolution initial conditions the simulated galaxies will have very different assembly histories.

\subsection{Hold-back}

In section 3.1 we presented evidence that hold-back of the grids introduces excess power into the small scale density field. Figure \ref{Fig:SimresPkz0a} demonstrated that this effect is only true for I7F9 and I7F10, and breaks down after I7F12, whereupon star formation physics reduces the formation of smaller structures. When the simulation resolution is low, the no-hold-back case shows less small-scale power than the  hold-back case. However, as the simulation resolution increases, the impact of holding back the grids is mitigated by star formation. The effect is probably still relevant on smaller scales, but on the scales of interest in this paper the no-hold-back and held-back cases show only slight differences in I7F12. 

From I7F15 the difference between the non-held back simulation and the non-held-back (no stars formed) simulation of I7F15 is -0.15. We know that, by this point, the effect of the additional grids on the power spectrum is not as important on the scales of 62.5 $h^{-1}$kpc which we are probing. This is not to say that there are not further differences on smaller scales but at this point the effect of the subgrid physics is expected to be dominant. 

\begin{figure*}
    \centering
    
     \begin{tabular}{cc}
     
     \includegraphics[scale=.5]{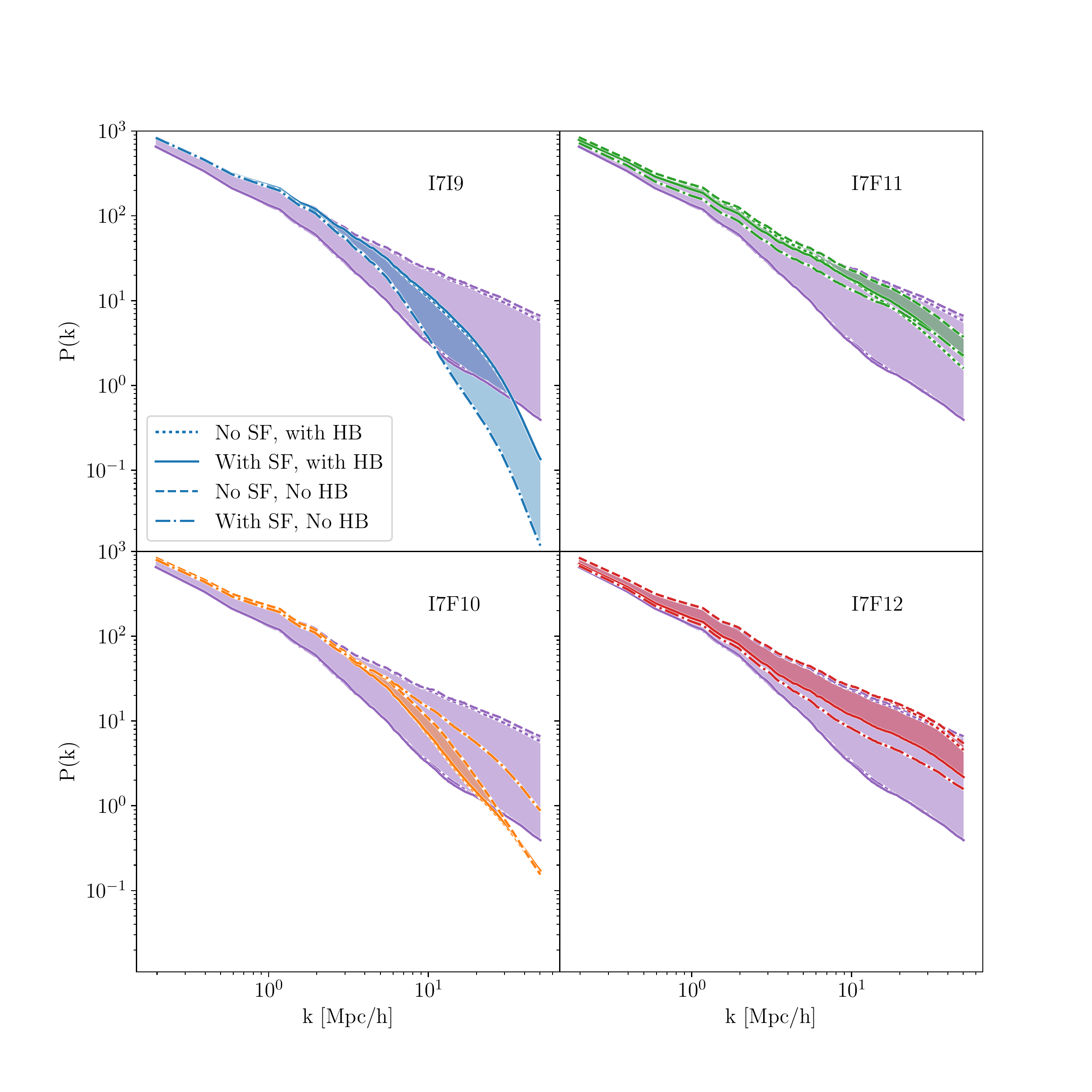} \\
     \includegraphics[scale=.5]{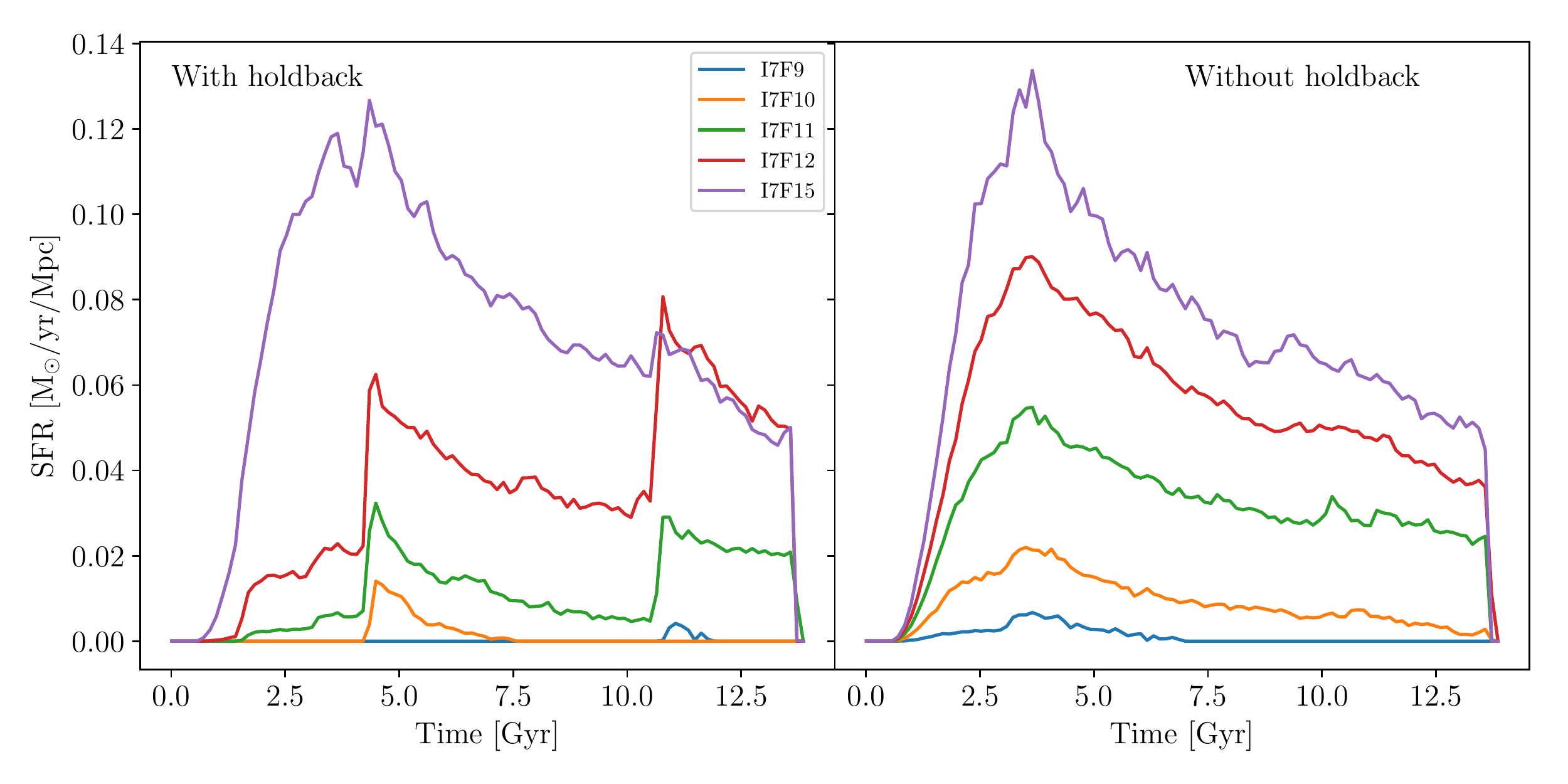}\\    
     \end{tabular} 
     \caption{The effect of star formation and the grid hold-back criterion at different simulation resolutions. Each of the four upper panels show the power spectra of eight different simulations. In each panel the purple lines, and region, show the power spectra of three simulations with an IC and simulation resolution of I7F15. These are the same in each panel. Also shown are the power spectra of simulations with the same IC resolution and different simulation resolutions. The blue, orange, green and red lines show the profiles for I7F9, I7F10, I7F11 and I7F12. The solid lines show the power spectrum where stars are allowed to form, and the AMR grids are held back; the dashed line shows the non- star-formation run with grids able to refine freely; the dotted line has no star formation but holds back the upper grid levels  and the dot-dashed line shows the evolution when stars form but the grid hold back is turned off. In the lowest resolution simulation, were very few stars form, the dominant effect is whether the grids are held back or not. This changes the power spectrum on small-scales by an order of magnitude. As the resolution increases, more stars form (bottom panels), and the effect of the grid hold-back effects smaller and smaller scales, the dominant effect is the star formation physics. The ICs are the same as used previously and is the green line in Fig. 1. The coloured region shows the difference between the star forming and non-star forming free refinement simulations. }
     \label{Fig:EvolL6L15}    
\end{figure*}

Figure \ref{Fig:EvolL6L15} illustrates the impact of simulation resolution on the recovered power spectrum a fixed IC resolution as well as the impact of the grid-hold-back implementation on scales greater than 62.5 $h^{-1}$kpc, as previously shown in Fig. \ref{Fig:SimresPkz0a}. However, we complement these simulations with an additional set. Here, we allow the grid-hold back, but suppress star formation by increasing the density threshold required to form stars to a high enough level that star formation becomes impossible using \texttt{n\_star} which has the side effect of turning off polytropic support. We also include a set of runs where the hold back is turned off but stars are allowed to form as before. This may introduce spurious numerical effects at early times due to the high resolution. 

In the low resolution I7F9 run, the star formation has minimal impact on the power spectrum, as almost no stars are formed, the difference in power is dominated by the refinement approach (i.e. whether the grids are held back or not). In the top left panel of Fig. \ref{Fig:EvolL6L15}, we see that the two simulation with hold-back are similar, while the no-hold-back simulation shows less small-scale power than before. 

In I7F12 the star formation is considerable (bottom left panel) and the power spectrum in the  grid hold-back case strongly diverges from the  non-held-back runs while the two non-held-back runs show similar power spectra regardless of whether the grids are held back or not. 

However, the grid hold-back tends to affect the power spectrum close to the resolution of those grid cells. In I9F11 there is no visible impact on the power spectrum in grid-hold-back run but this may be occurring on much smaller scales than we are probing. Each grid release is l$_{\mathrm{max}}$-n, where $n$ is the number of levels held back. So at I7F15 level 15 is released at a$_{\mathrm{exp}}$=0.8 while for I7F13 it is level 13 which is released at that time. Lower levels are increasingly less affected and, therefore, to examine the importance of the grid-hold back on simulations with higher simulation resolution we must use other methods. For example, the cosmic star formation rates shown in the lower-most panel of the Fig. \ref{Fig:EvolL6L15} demonstrate that the star formation rate is still strongly affected by the grid-hold-back, with an offset at the release of the highest grid level. 

When we allow star formation but prevent grid hold-back the difference between the two star forming cases is very small ($\delta^{\mathrm{SF,holdback}}_{\mathrm{SGF,no holdback}} \sim$0.06) in I7F15, but an order of magnitude larger in I7F12. This is due, in part, to the increase in star formation. With grid hold-back in I7F15 there is only a small difference in the amount of stars formed, but in I7F12 almost 1.5 times as many stars are formed in the non-held-back case, lowering the small-scale power.

\subsection{Number of levels and the force resolution}
 To further examine the effect of the resolution and the number of refinement levels on the  power spectrum at $z=$0 we carry out an additional set of simulations. We fix the physical force resolution at $z$=0 {\it and} the initial conditions and change the box size. This means that the initial conditions are not the same as in the other runs, but the same random number seed was used. 

\begin{figure*}
    \centering
    
     \begin{tabular}{cc}
     \includegraphics[scale=.62,trim={.5cm .5cm .5cm 0cm},clip]{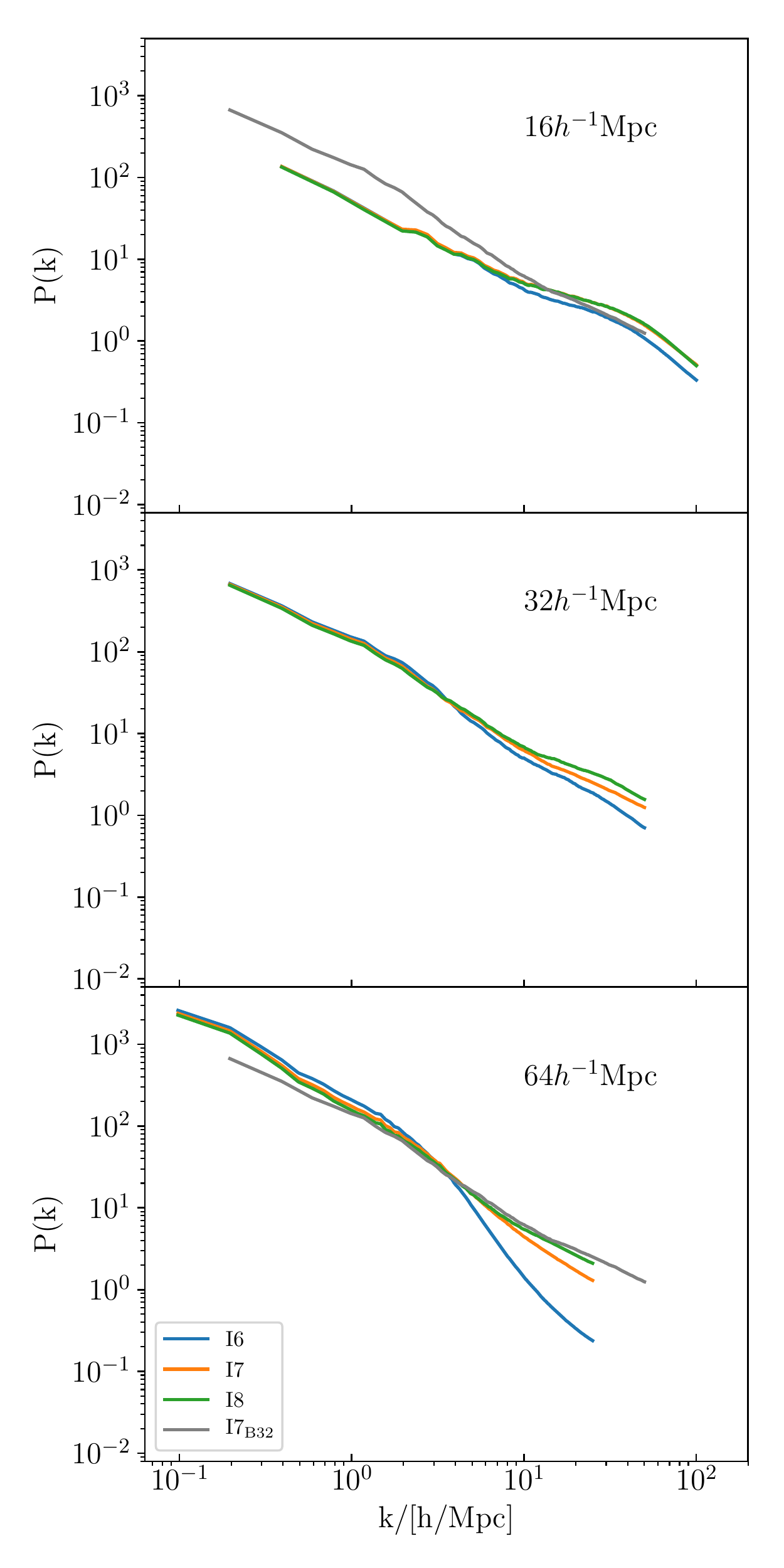}&
     \includegraphics[scale=.62,trim={.5cm .5cm .5cm 0cm},clip]{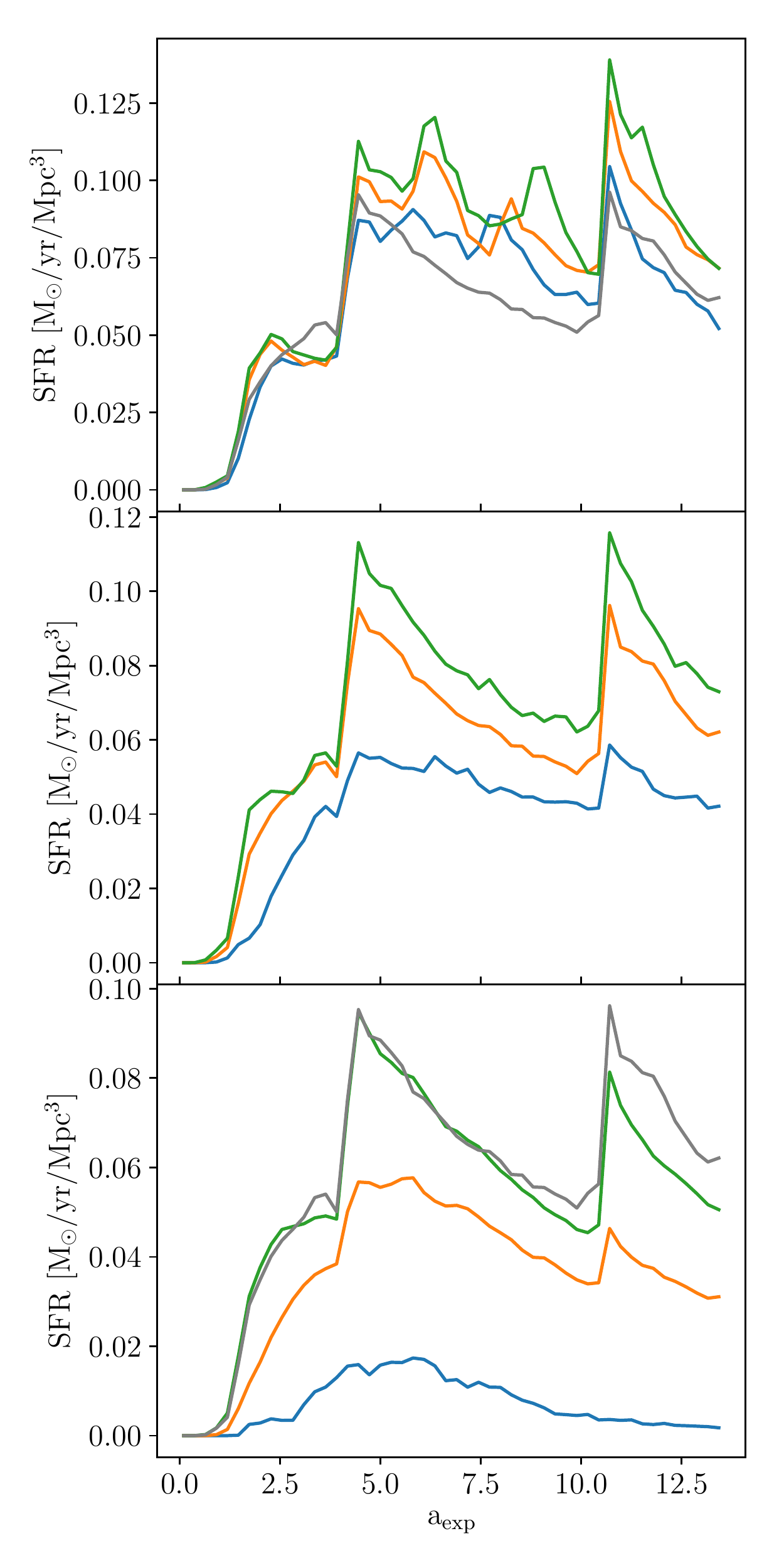}\\    
     \end{tabular} 
     \caption{Left panel: The gas power spectra in different simulation volumes. The meaning of each line colour is given in the legend and the box size is given in each panel. The grey line is the same in each panel and indicates the I7I13 simulation in the 32 $h^{-1}$Mpc volume. Right panel: The corresponding star formation rates.}
     \label{Fig:Diffboxes}    
\end{figure*}

In Fig. \ref{Fig:Diffboxes} we show the power spectra of I6F12, I7F12 and I8F12, in a 16 $h^{-1}$Mpc box, I6F13, I7F13 and I8F13 32 $h^{-1}$Mpc box and  I6F14, I7F14 and I8F14 in a 64 $h^{-1}$Mpc box. In the smaller boxes the large-scale power is less than in the 64 $h^{-1}$Mpc box. We include star formation here. These are not as easy to compare as the previous sections because the initial conditions are not identical (e.g. the same random number seed is used but a different region of the primordial power spectrum is being probed). Nevertheless, we see that the power spectra in the smallest box converges, with only a small offset between I7F12 and I8F12 ($\delta \sim$0.05). This may indicate that the initial conditions must have a resolution of between 0.125 $h^{-1}$Mpc and 0.0625  h$^{-1}$Mpc to correctly model the evolution of the power spectrum down to 0.03125 $h^{-1}$Mpc. Furthermore, the three 16 $h^{-1}$Mpc box simulations all produce similar amounts of stars. The total star formation increases with resolution as before although not to the same degree (for the I6, I7, and I8 runs the stellar mass ratios are 1: 1.16, 1.26 for the  16 $h^{-1}$Mpc box, 1:1.48:1.73 for the 32 $h^{-1}$Mpc box and 1: 5.12: 7.66 for the 64 $h^{-1}$Mpc box). This explains the increasing divergence of the gas power spectrum in the different simulation volumes, as discussed previously. 

However, I6F12, I7L13, and I8L14 have the same force and initial conditions resolution, the only difference is the box size and the subsequent evolution of the material. In the different boxes, $\delta^{\mathrm{I6F12}}_{\mathrm{I8I14}}$ ranges from -0.75 (at low $k$) to 0.07 (at high $k$) while $\delta^{\mathrm{I7F13}}_{\mathrm{I8I14}}$ ranges from  -0.3 to 0.2 and low and high $k$, respectively. At the same time the star formation histories of these three simulations similar with 7\% and 12\% more stars produced in I6F12 and I7I13 relative to I8I14.

 It is noteworthy that when we tested a simulation with a resolution of 1 kpc there where still peaks/offsets in the star formation history corresponding to the grid release epochs using the public code \footnote{Albeit using a version of \textsc{RAMSES} downloaded after the one used for the majority of the paper, and with different initial conditions/cosmology.}. Although for much of this paper we have a maximum spatial resolution which is more coarse than 1 kpc, \citep[I8F15 has a resolution of 1$h^{-1}$kpc, comparable to the gravitational softening length used Illustris-2 simulation for gas or Illustris-1 for dark matter particles, see][for details]{Vogelsberger2014} the same feature is still visible at resolutions where the galaxy discs are resolved. It implies that this is a typical result, and not due to spurious effects caused by the relatively low resolution adopted here. This demonstrates that the shock to the system is due to the properties of the gas, and not the star formation physics. In other words, more sophisticated treatment of star formation beyond the Kennicutt-Schmidt relation adopted in the public version of \textsc{RAMSES} may resolve the star formation rate issue. However, because it has an impact on the underlying gas distribution as well (even if this is second order if the star formation rate is high), there may be spurious numerical effects due to additional clumping in the gas.

\section{A solution to the problem}
\label{Sec:New}
 As grid release impacts the power spectrum and the star formation rate even at 1 kpc resolution, we need to reduce or mitigate the problem. The simplest way would be to remove the grid hold back entirely, and allow the system to evolve with a constant co-moving resolution. This is not preferred because: (1) much of the subgrid physics assumes approximately fixed physical resolution; (2) the resulting evolution of gas properties might be due to the numerical effects of the changing physical resolution rather than physics; (3) at early times the force resolution will be high compared to the mass of dark matter and stars, potentially introducing two-point scatter in the N-body component. 

In the following section we concentrate on simulations with high resolution, where the power spectrum is dominated by the star formation effects. As there is a signature of the grid hold back in the star formation rate we will concentrate on the cosmic star formation history as the principal diagnostic of the effectiveness of our new method. We are interested in the flow of gas on galaxy, or sub-galactic, scales, and so the power spectrum is less effective an approach, due to the limited resolution of the grid cells used to calculate the Fourier Transform. 

An alternative solution to the global refinement approach to the grid hold back of \citep{Dubois2014} is to maintain the grid-hold back, but instead of releasing the grids via a on/off switch we transition more gradually. The prescription of grid release used in our modified version of \textsc{RAMSES} uses a Logistic function of the form:

\begin{equation}
n_{\rm refine}(a) = n_{\rm max}+(n_{\rm max}-n_{\rm f})\left(1-\frac{1}{1+\exp^{-S(a-c)}}\right),
\label{Eqn:newref}
\end{equation} 
where $S$ sets the steepness of the slope, $c$ is the epoch at which the refinement level is released, $n_{\rm refine}$ is the current number of particles in a cell required for refinement under the quasi-Lagrangian approach, and  $n_{\rm final}$ is the final number of particles required for refinement, usually 8 \citep{Dubois2014}.

We take $c$ from the current \textsc{RAMSES} code,

\begin{equation}
c=4^{1/n_d}2^{-(l_{max}-l_{i})},
\end{equation}
\label{Eqn:cc}
where $n_d$ is the number of dimensions, $l_{max}$ is the maximum refinement level, and $l_i$ is the {\it i}th refinement level. Using these equations the grid release occurs more gradually, with a rate that depends on the value of $S$. 

\begin{figure}
    \centering
     \begin{tabular}{c}
     \includegraphics[scale=.55,trim={.4cm .5cm .5cm 0cm},clip]{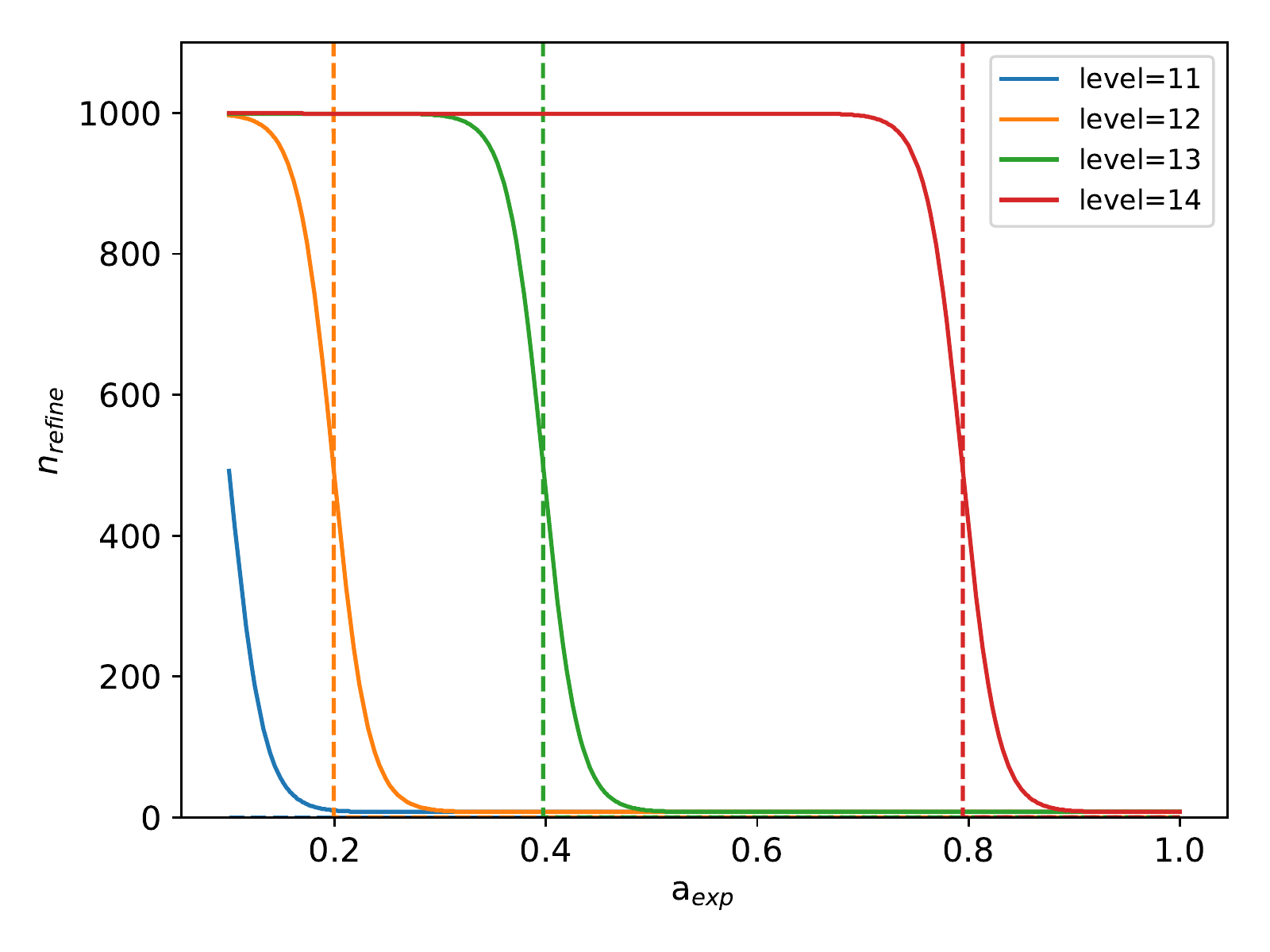}\\
     \includegraphics[scale=.55,trim={.4cm .0cm .5cm 0cm},clip]{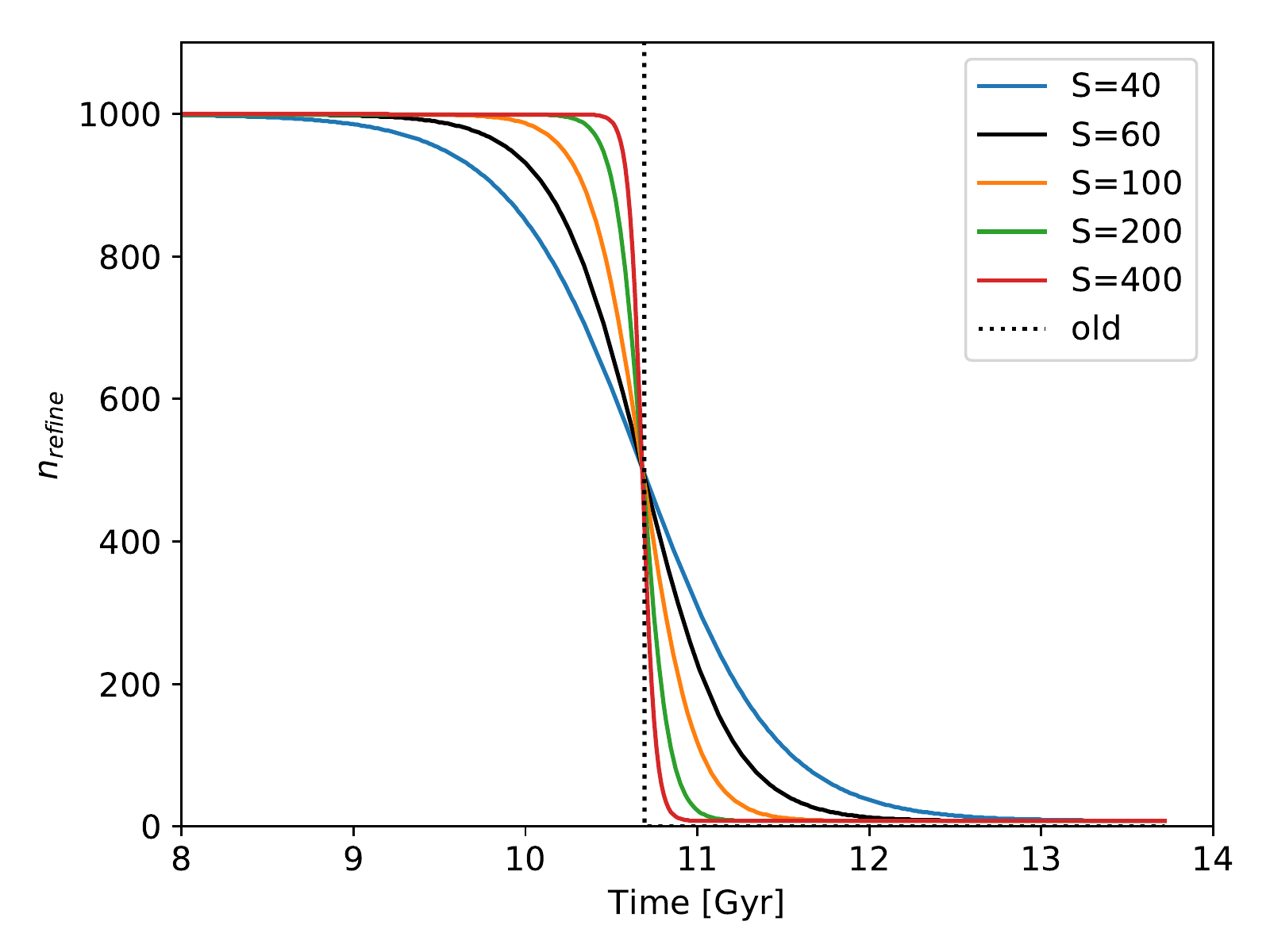}\\     
     \end{tabular} 
     \caption{Upper panel: The evolution of $n_{refine}$ from Eqn. \ref{Eqn:newref} using the new method (solid lines). The only grid release epochs are marked as vertical dashed lines. Lower panel: the rate in Gyr at which \texttt{m\_refine} falls for different values of $S$.}
     \label{Fig:functionrosdahl}    
\end{figure}

The upper panel of Fig. \ref{Fig:functionrosdahl} shows how this new refinement method evolves. The dashed line shows a depiction of the former method, where the upper refinement levels are forbidden until certain values of a$_{exp}$. In terms of the { \texttt{m\_refine}} criterion, the number of particles per cell is effectively infinite before the transition and the user provided value (usually 8) afterwards. In the new method, given by Eqn. \ref{Eqn:newref}, before the transition \texttt{m\_refine} is set to be very high (for example 1000 particles/cell). The number of particles required to refine drops over time to reach the user defined value shortly afterwards. The lower panel shows how the rate of the fall is affected by the value of the parameter $S$. At very high $S$ the slope is effectively the same as the old criterion.  

During the transition period, the cells begin to refine according to Eqn. \ref{Eqn:newref}. Over the transition period high resolution regions can be said to have higher resolution than lower density regions, but this is entirely within keeping with the AMR paradigm. High density regions refine first, when the value of $n_{\rm refine}$ is still high, while lower density regions (but with more than $n_{\rm f}$ particles per cell) refine later. This follows the natural refinement method of AMR codes, where the accumulation of matter triggers refinements and so should not introduce unphysical effects.  

\begin{figure*}
    \centering
    
     \begin{tabular}{ccc}
     \includegraphics[scale=.8,trim={3.5cm .5cm 3.5cm 0cm},clip]{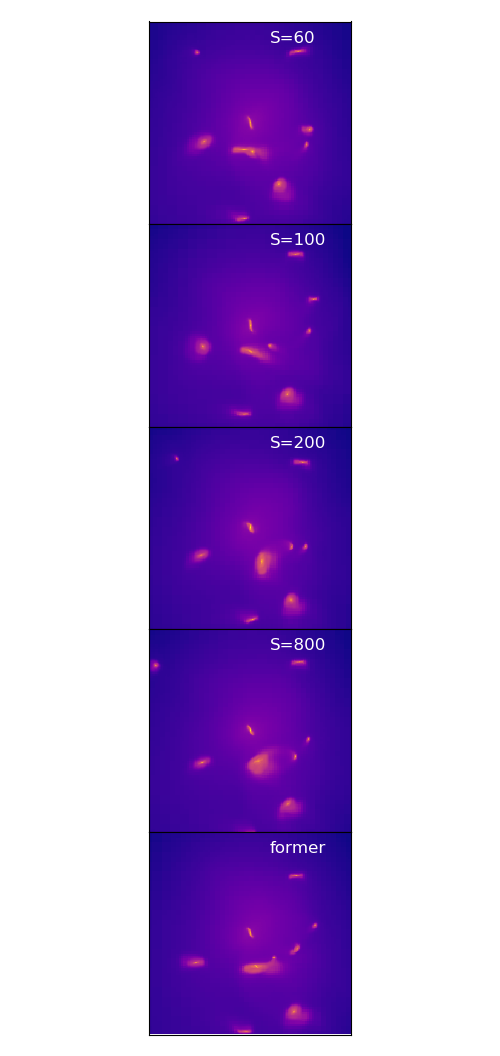} &
      \includegraphics[scale=.8,trim={3.5cm .5cm 3.5cm 0cm},clip]{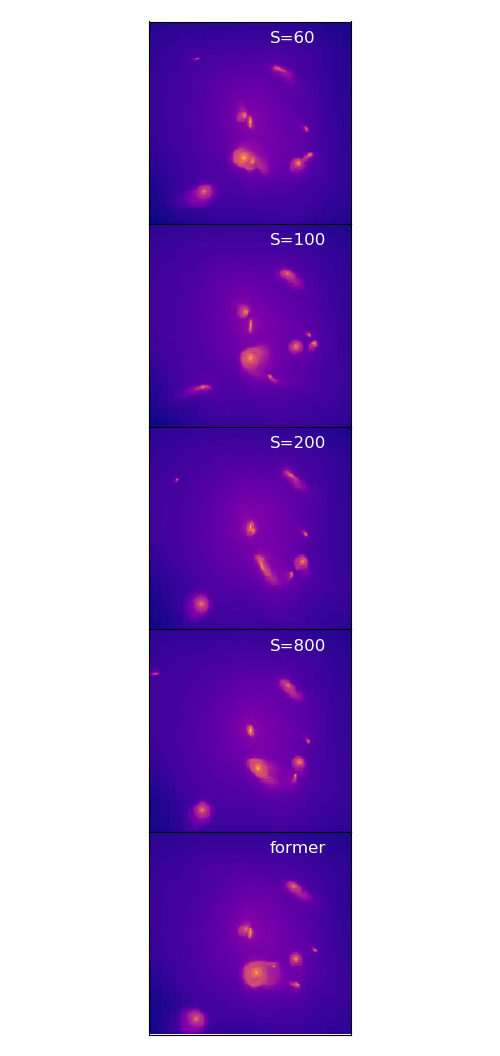}   &
      \includegraphics[scale=.8,trim={3.5cm .5cm 3.5cm 0cm},clip]{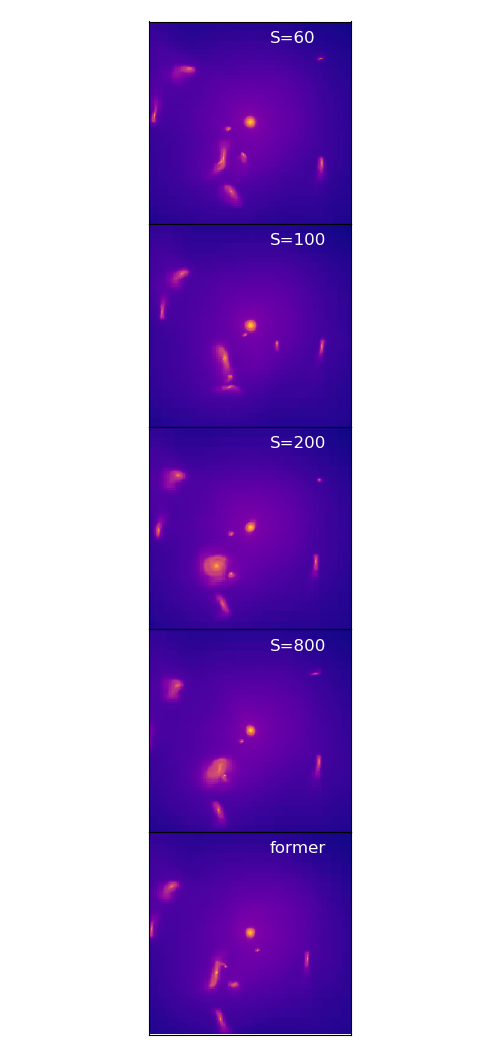}   \\   
      x & y & z \\    
     \end{tabular}     
     \caption{Gas density averaged along the line of sight for the inner 68\% of the virial radius along each principal direction for the I7F15 simulation. Each row is for a diven value of $S$ from Eqn. \ref{Eqn:newref}. Gas is coloured from purple (low density), to yellow (high density). The equivalent for I8F15 is given in Fig. \ref{Fig:Clusterrosdahlo2}}
     \label{Fig:Clusterrosdahlo1}    
\end{figure*}

Figure \ref{Fig:Clusterrosdahlo1} shows the most massive halo for a number of test simulations. The figure shows the halo at $z=0$ in a I7F14 simulation, with each panel showing a different set of parameters in Eqn. \ref{Eqn:cc}. The top-left panel shows the simulation carried out using the {\it most current version of \textsc{RAMSES} and using the original global refinement method}. The values of the slope in Eqn. \ref{Eqn:newref} are shown as $S$ in each panel. Although there are some differences, particularly in the angular momentum of some of the discs, the overall appearance of the gas discs seen are not hugely dissimilar, with some simulation-to-simulation differences. For example, panel $S=200$ the $y$-plane is more different to the former method panel than any of the other runs, at least by eye. There are differences in the precise inclination or position of galaxies in each panel, due to the essentially chaotic behaviour of cosmological systems. Globally, however, there is not a strong systematic difference in morphology, for example.  This means that our new method does not strongly affect galaxy properties, although it does appear to impact galaxy positions. Having said this, the initial conditions of these simulations are a factor of two lower resolution than the highest resolution simulation discussed previously, so the galaxies may not be completely resolved. 

In Fig. \ref{Fig:Clusterrosdahlo2} we compare the $S=60$ simulation with a simulation carried out without the new method in the I8F15 simulation with a resolution of 1$h^{-1}$kpc, and again, the overall shape of the galaxies is not vastly different. If there is any effect of the new method it is that galaxies appear more relaxed than using the original approach.

\begin{figure}
    \centering
     \includegraphics[scale=.6,trim={.5cm 1.5cm .5cm 2cm},clip]{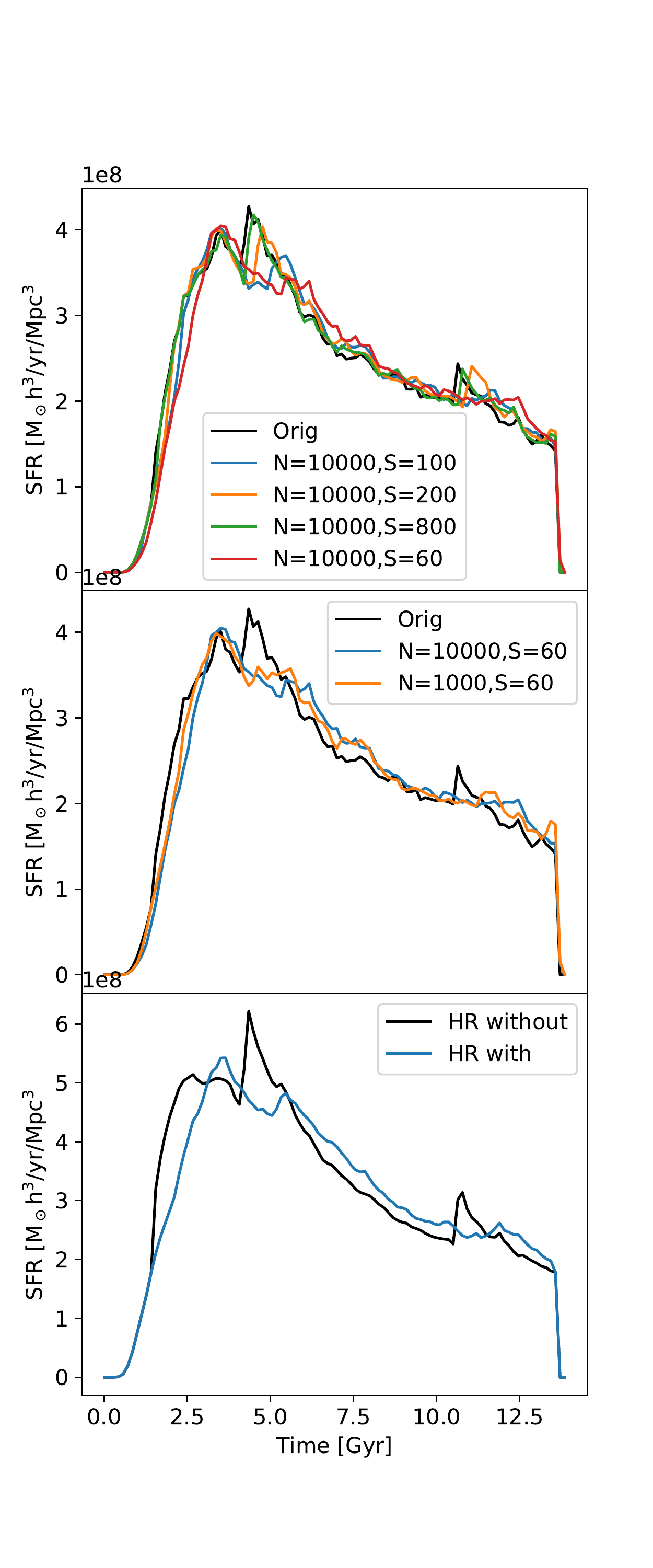}
     \caption{Top panel: The star formation rates for the different test simulations using the new method for maximum \texttt{m\_refine}=10000 (blue, orange, green and red lines) compared to the original grid release approach (black line). Middle panel: The effect of the number of particles before the transition ($N$). Bottom panel: For the I8F15 run using $N$=10000, $k$=60.}
     \label{Fig:SFHRosdahl}    
\end{figure}

However, one of the primary goals of this method was to remove the spurious peaks in the star formation history caused by the global refinements. Figure \ref{Fig:SFHRosdahl} shows the original SFH for the I7F14 (and I8F15) simulation in black. There are strong peaks at around 5 and 11 Gyr corresponding to the global refinements. The size of the peak is softened with decreasing $S$, until they have almost vanished at $S$=60 for both I7F14 and I8F15. The \texttt{m\_refine} value, however, does not have a strong impact so long as it is sufficiently high to prevent refinements too early.  

In order to check the impact of the new refinement approach on the outcome of simulations we explored the properties of halos from each of the test simulations. Halos were identified using a halo and subhalo finder called AHF \citep[Amiga Halo Finder,][]{Knollmann2011,Gill2004} which uses an AMR approach to identify matter bound to density peaks. 

\begin{figure*}
    \centering
     \includegraphics[scale=0.85]{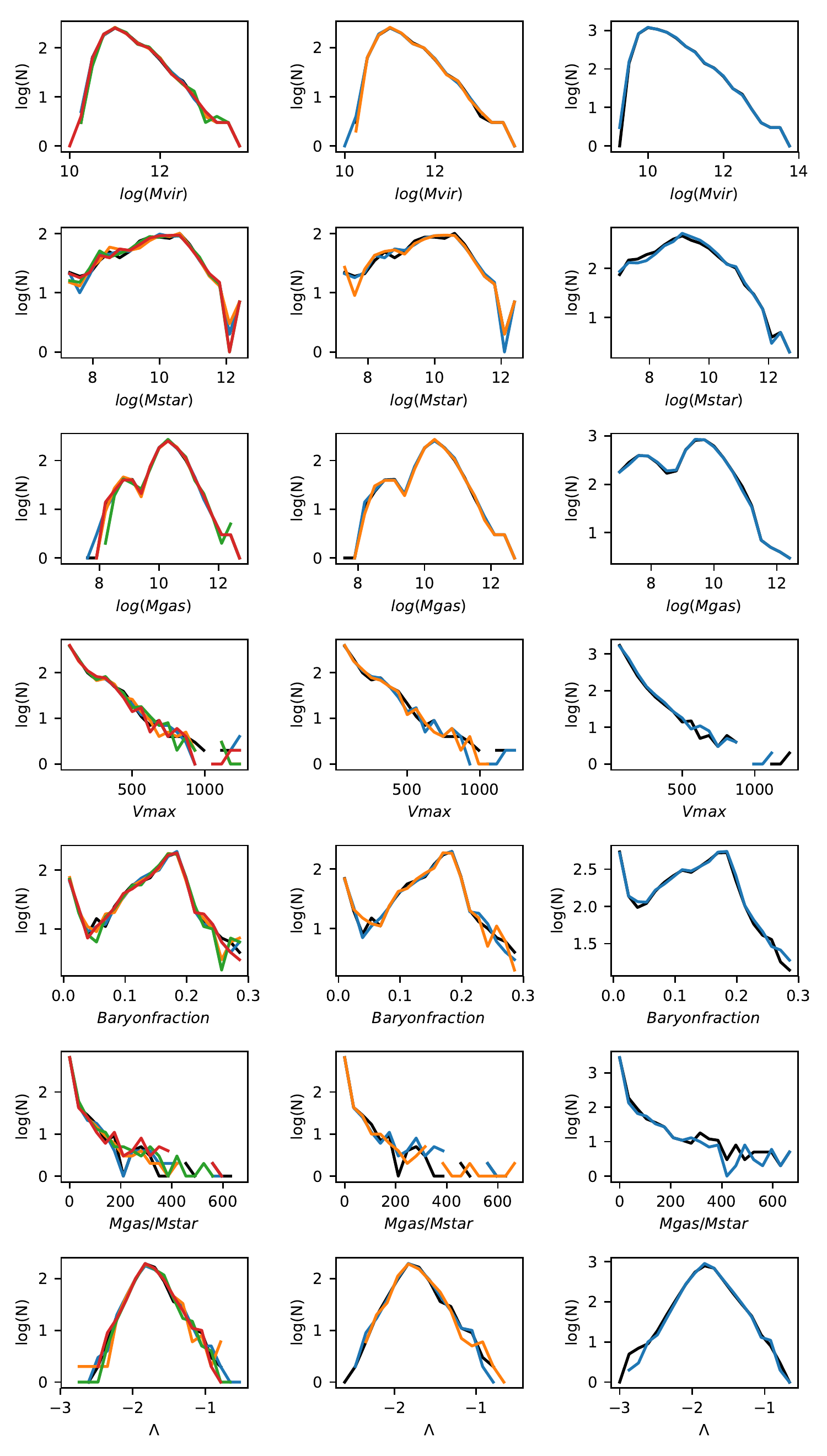}
     \caption{Various galaxy properties in the different simulation coloured in the same way as in Fig. \ref{Fig:SFHRosdahl}. $\Lambda$ is the halo spin. }
     \label{Fig:HalostatsRosdahl}    
\end{figure*}

Fig. \ref{Fig:HalostatsRosdahl} illustrates that overall the halo properties are unaffected by the new criterion, and that overall the resolution has, by far, the dominant affect on the distribution, particularly the gas mass of halos. 

Within the regime of interest in this paper the new refinement approach seems to produce considerable fewer artefacts in the star formation history, but otherwise does not strongly effect the properties of halos. 

This method does not completely remove stepwise effects. For example, the I7F12 simulation still shows a sudden change in star formation rate as the upper grid level is released, although the change in effective star formation rate efficiency is more gradual. In the case of a full simulation up to high resolution (e.g. F7F15), the peaks are invisible. This method appears to alleviate most of the spurious impact of the global refinements on the star formation.

\section{Discussion \& Conclusions}
\label{Sec:Conclusions}

We have examined the properties of the gas power spectrum in simulations using the \textsc{RAMSES} adaptive-mesh-refinement code. To this end we produced a series of simulations with different initial condition resolutions and simulation force resolutions. We calculate the power spectrum down to a scale of 62.5 $h^{-1}$kpc. We have studied how the different resolutions impact the power spectrum evolution in \textsc{RAMSES}, and noted the impact of the various choices made in the code implementation. 

We found that for the lower resolution simulation runs (IxF9) the choice of whether the simulation was run with static grids or AMR had a greater impact on the gas power spectrum, by an order of magnitude, than the difference between varying IC resolutions with AMR. This is not true for the dark matter, which shows a steady convergence to the fixed grid run with increasing IC resolution. 

This difference is a result of the grid-hold-back criterion in \textsc{RAMSES}. In order to maintain the physical force resolution of the ISM, the upper grid levels must be held back so that the volume can be refined each time the simulation volume doubles in physical size. However, as these levels are released, and the volume refined, there is a sharp increase in the small scale power of the AMR simulations. This leads to a considerable excess of power at high $k$. This hold-back procedure has the effect of limiting the maximum force resolution of the simulation at early times. However, a constant force resolution is needed to avoid apparent evolution of ISM properties due to changing physical resolution \citep[Dubois: private communication]{Dubois2012}. 

As the force resolution of the simulation increases we can expect the increase of small scale power to be important on smaller and smaller scales, and on smaller scales the star formation physics is more important than the effect of grid refinement choice. The influence of the grid hold-back can also be seen in the star formation history as sudden offsets in the star formation rate. 

 It is to be expected, however, that the precise differences in the power spectrum with the star formation physics depends on the particular properties of the implemented physics \citep[e.g.][]{vanDaalen2011}. 

The grid hold-back has a considerable impact on the star formation history, which probes scales smaller than the scale on which we can measure the power spectrum ($k>$62.5 $h^{-1}$kpc). The hold-back results in large changes in the star formation efficiency when the grids are released. However, this effect is reduced as the resolution increases. This is potentially due to the highest grid levels occupying smaller and smaller regions of the simulation volume. For example, at higher levels the refinements take place throughout a galaxy, whereas at higher levels only the very depths of the potential wells are refined. Thus, once the resolution is high enough the effect of the grid-hold back on the star formation is reduced. This can be clearly seen in the lower panel of Fig. \ref{Fig:EvolL6L15}. Once the feedback effect is considerable, the effect of the grid-hold-back is much less important.

There is no sign of convergence in the star forming cases as we expand to higher simulation and initial conditions resolution. This suggests that higher resolution or a more accurate representation of the physics is required. Our result emphasises the importance of the subgrid physics, which is the most critical aspects of hydrodynamical simulations. \citet{Scannapieco2012} note that on a galaxy scale the implementation of the feedback is more important the the numerical methods used in the code (e.g. treecode, SPH, AMR, moving mesh, etc). Thus, parameter tuning is essential. For low resolution simulations with few grids and low star formation the grid-hold-back effect is influential on the results of the gas. However, as the number of stars increases, the subgrid physics has a much stronger impact on the simulation outcomes. It is difficult, however, to reach the total stellar mass produced in the I6F15 simulation, for example, to reach the total stellar mass of the I7F15, even if the star formation efficiency is increased from 0.02 to 0.2. 

Figure \ref{Fig:image} demonstrates the importance of the initial conditions/mass resolution at defining the distribution of substructure even at later times \citep[see ][ for a further discussion]{Schneider2016}. Thus, knowing the initial conditions resolution is vital to interpreting simulations because the galaxy assembly history of objects in I6F15 and I7F15 are very different. 

 We present an alternative approach to the grid-hold-back in section \ref{Sec:New}, which reduces the impact of the grid release on the star formation history. This is important, because much of the baryonic behaviour depends on the star formation, such as feedback and chemical evolution. By introducing a step-wise artefact, like the ones shown in this paper, one of the fundamental strengths of simulations is undermined -- the ability to explore the time evolution of objects. With our new method, which softens the grid-release transition, we are able to give the system chance to adapt to the new resolution, and so it is not so strongly shocked. Our method does not strongly affect the bulk properties of galaxies in the resulting simulation. However, the strength of the new approach is that it removed the offsets in the star formation history which occur at a$_{exp}$=0.2, 0.4 and 0.8. These have been smoothed away, especially for the $S=$60 simulation, shown in Figure \ref{Fig:SFHRosdahl}. Although the bulk properties have not been strongly affected there is an impact on the detailed properties of the galaxy, particularly in terms of its evolution. The old method introduced strong time dependent artefacts in the star formation properties of galaxies, which will impact the distribution of gas, the chemical enrichment, luminosity, etc. at those times. With the new method, these artefacts have been strongly reduced, and so following the properties of objects through time will no longer have the same sudden changes.

\section*{Acknowledgements}
The authors thank the Korea Institute for Advanced Study for providing computing resources (KIAS Center for Advanced Computation Linux Cluster System) for this work. This work was supported by the Supercomputing Center/ Korea Institute of Science and Technology Information, with supercomputing resources (KSC-2016-C3-0071), and the simulation data were transferred through a high-speed network provided by KREONET/ GLORIAD. JR acknowledges support from the ORAGE project from the Agence Nationale de la Recherche under grant ANR-14-CE33-0016-03. ONS also thanks Yohan Dubois, Romain Teyssier, Christophe Pichon, Julien Devrient and Brad Gibson for interesting and helpful discussions on this subject. In this paper we made use of the \textsc{numpy} \citep{Walt2011}, \textsc{scipy} \citep{Jones2001}, \textsc{matplotlib} \citep{Hunter2007} python libraries and \textsc{IPython} \citep{Perez2007}. This research made use of Astropy, a community-developed core Python package for Astronomy \citep{Astropy2013}. The authors thank the referee for their comments, which improved the paper.

\bibliographystyle{mnras} 
\bibliography{bibliography}

\appendix
\section{Resolution issues}

Fig. \ref{Fig:DMonlylev} demonstrates that the natural evolution of the dark matter resolution depends on the initial conditions in a non-trivial way, due to the clusterning of the dark matter under gravity. The hydrodynamic component is able to cluster at much higher densities, allowing us to achieve F15 for I6, when the dark matter is only able to reach level 12 on its own. This means that there is a risk of numerical effects in the hydrodynamical simulations due to the interactions of particles. We use the final values achieved by the I6 and I7 dark matter simulations in Fig. \ref{Fig:DMonlylev} to set the maximum resolution of a set of hydrodynamical simulations. This was done to test whether the numerical effects caused by high resolution affect the conclusions of this paper. 

\begin{figure}
    \centering
     \includegraphics[scale=.5]{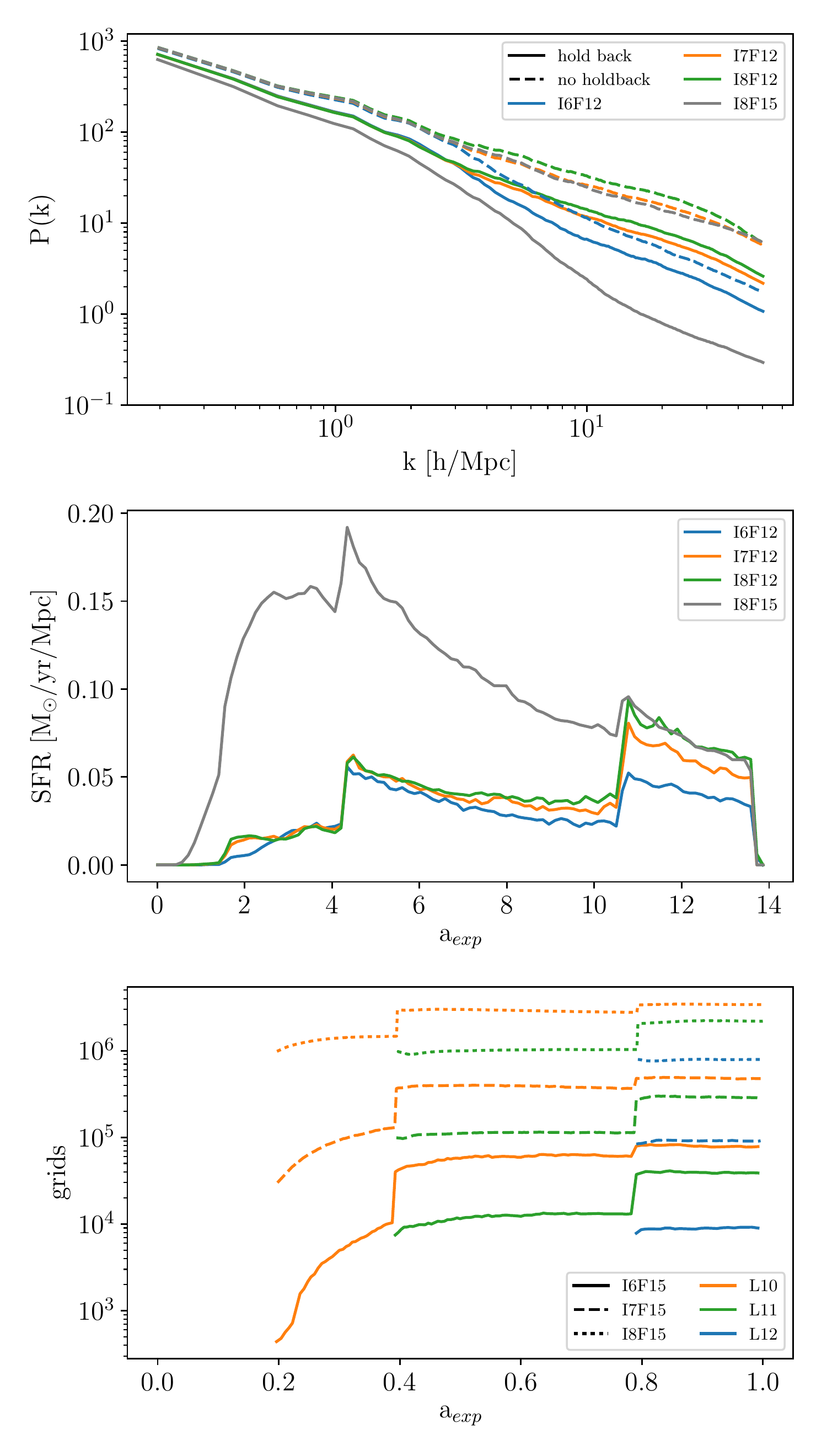}
     \caption{The same plots as in Fig. \ref{Fig:SimresPkz0} but for IxF12. F12 is the natural resolution of the I6 dark matter only simultion. Note that there is the drop in power in the star forming power spectrum, as expected from Fig. \ref{Fig:EvolL6L15} but that the star formation rate is strongly affected by the global refinements. The grey lines are the I8F15 line from Fig. \ref{Fig:SimresPkz0} for comparison.}
     \label{Fig:Restest12}    
\end{figure} 

\begin{figure}
    \centering
     \includegraphics[scale=.5]{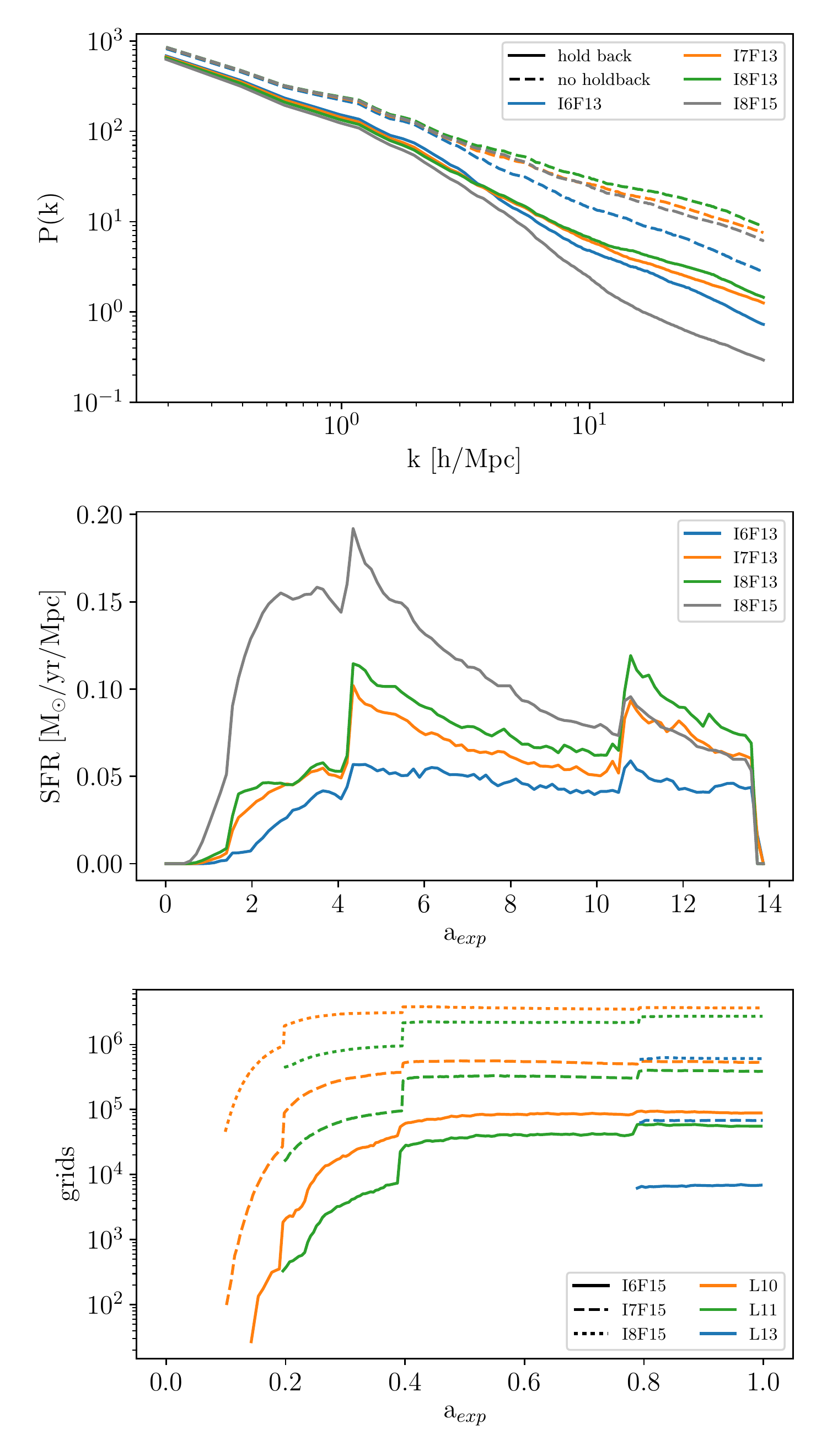}
     \caption{The same plots as in Fig. \ref{Fig:SimresPkz0} but for IxF13. F13 is the natural resolution of the I7 dark matter only simultion.}
     \label{Fig:Restest13}    
\end{figure}

Figs. \ref{Fig:Restest12} and \ref{Fig:Restest13} show that even accounting for numerical effects caused by the highest simulation resolution runs, the system still undergoes stepwise changes in the star formation rate when global refinements are released.

\begin{figure*}
    \centering
    
     \begin{tabular}{c}
      \includegraphics[scale=.3,trim={.5cm .5cm .5cm 0cm},clip]{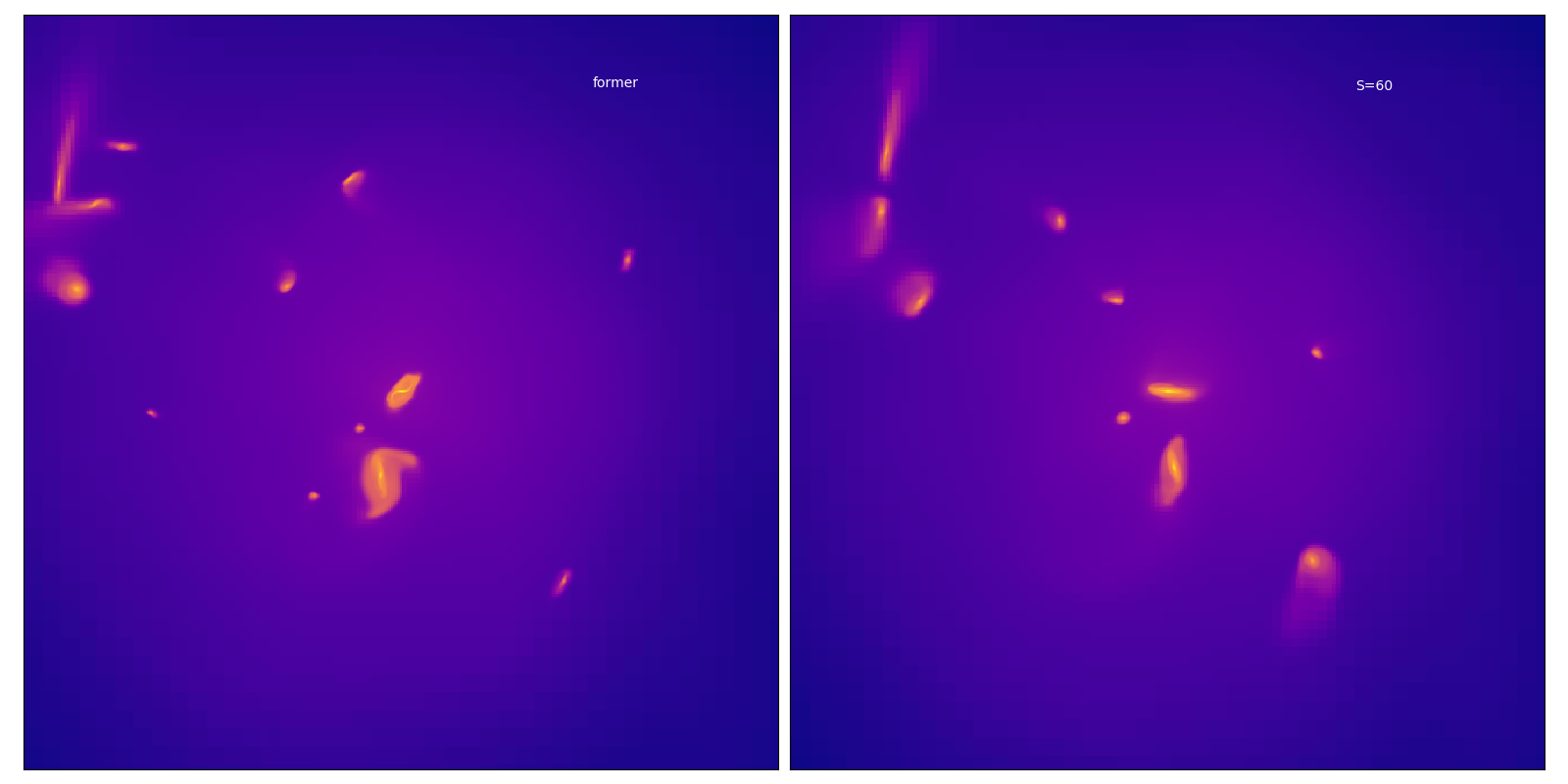}   \\      
     \end{tabular}     
     \caption{Comparison between the former global refinement method (left) and the new method with $S=60$ (right) for I8F15. Gas is coloured as in Fig. \ref{Fig:Clusterrosdahlo1}} 
     \label{Fig:Clusterrosdahlo2}    
\end{figure*}

\end{document}